\documentclass[12pt,longbibliography]{article}
\usepackage{amsmath,amssymb,amsthm,amsxtra,overpic,bbm,bm,epsfig,subfigure}
\usepackage{booktabs}
\usepackage{hyperref}
\usepackage{mathrsfs}
\usepackage{graphicx}
\usepackage{multirow}
\usepackage{makecell}
\usepackage{verbatim}
\usepackage{color}
\usepackage{comment}
\usepackage{epstopdf}
\usepackage{float}
\usepackage{amsmath}
\usepackage{amssymb}
\usepackage{tikz}
\usepackage{wrapfig}
\usepackage{tikz-feynman}
\usepackage{verbatim}
\numberwithin{equation}{section}
\usepackage{cite}
\usepackage{hyperref}
\usepackage{url}
\usepackage{slashed,stmaryrd}

\newcommand*{\dif}{\mathop{}\!\mathrm{d}}

\begin{document}
	\vspace{0.2cm}
\begin{center}
	{\Large\bf Constraining Light Mediators via Detection of Coherent Elastic Solar Neutrino Nucleus Scattering}
\end{center}
\vspace{0.2cm}

\begin{center}
	{\bf Yu-Feng Li}~$^{a,~b}$~\footnote{E-mail: liyufeng@ihep.ac.cn}
	{\bf Shuo-yu Xia}~$^{a,~b}$~\footnote{E-mail: xiashuoyu@ihep.ac.cn (corresponding author)},
	\quad
	\\
	\vspace{0.2cm}
	{\small
		$^a$Institute of High Energy Physics, Chinese Academy of Sciences, Beijing 100049, China\\
		$^b$School of Physical Sciences, University of Chinese Academy of Sciences, Beijing 100049, China}
\end{center}

\begin{abstract}
Dark matter (DM) direct detection experiments are entering the multiple-ton era and will be sensitive to the coherent elastic neutrino nucleus scattering (CE$\nu$NS) of solar neutrinos, enabling the possibility to explore contributions from new physics with light mediators at the low energy range. In this paper we consider light mediator models (scalar, vector and axial vector) and the corresponding contributions to the solar neutrino CE$\nu$NS process.
Motivated by the current status of new generation of DM direct detection experiments and the future plan, we study the sensitivity of light mediators in DM direct detection experiments of different nuclear targets and detector techniques.
The constraints from the latest $^8$B solar neutrino measurements of XENON-1T are also derived. Finally, We show that the solar neutrino CE$\nu$NS process can provide stringent limitation on the $ L_{\mu}-L_{\tau} $ model with the vector mediator mass below 100 MeV, covering the viable parameter space of the solution to the $ (g-2)_{\mu}$ anomaly.
\end{abstract}

\newpage

\section{Introduction}

The observation of coherent elastic neutrino-nucleus scattering~\cite{Freedman:1973yd,Freedman:1977xn} (CE$\nu$NS) at the COHERENT experiment in the spallation neutron source at Oak Ridge National Laboratory~\cite{COHERENT:2017ipa,COHERENT:2020iec,Akimov:2021dab} has opened a new and powerful tool on a variety of topics in the fields of high-energy physics, nuclear physics, astrophysics and cosmology. The CE$\nu$NS detection has not only provided a unique probe to the nuclear neutron density distributions~\cite{Cadeddu:2017etk,Cadeddu:2020lky,Cadeddu:2021ijh,Ciuffoli:2018qem,Papoulias:2019lfi,Coloma:2020nhf}, but also offered a precision test of the physics beyond the Standard Model (SM), including the weak mixing angle~\cite{Papoulias:2017qdn,Cadeddu:2018izq}, the neutrino electromagnetic properties~\cite{Cadeddu:2018dux,Cadeddu:2019eta,Kim:2021lun}, the nonstandard neutrino interactions~\cite{Coloma:2017ncl,Liao:2017uzy,Denton:2018xmq,AristizabalSierra:2018eqm,Giunti:2019xpr} and the light mediators~\cite{Farzan:2018gtr,Abdullah:2018ykz,Ge:2017mcq,Cadeddu:2020nbr}. On the other hand, the observation of the CE$\nu$NS process have important implications on the neutrino floor~\cite{Boehm:2018sux,Gonzalez-Garcia:2018dep,Papoulias:2018uzy,Chao:2019pyh,AristizabalSierra:2021kht,OHare:2020lva} in the Dark Matter (DM) direct detection and the observation of astrophysical neutrino fluxes from the supernovae~\cite{Lang:2016zhv,Pattavina:2020cqc,RES-NOVA:2021gqp,Raj:2019sci,Raj:2019wpy,Huang:2021enl}, the collapsing supermassive stars~\cite{Munoz:2021sad} and the primordial black holes~\cite{Calabrese:2021zfq}.
In addition to the observation at spallation neutron sources~\cite{Barbeau:2021exu,Baxter:2019mcx}, there are also intensive interests in the CE$\nu$NS detection from man-made reactor neutrinos~\cite{CONUS:2020skt,CONUS:2021dwh,CONNIE:2019xid,CONNIE:2021ngo}. Although no evidence of the reactor neutrino CE$\nu$NS process has been observed, there are already interesting limits on a variety of new physics scenarios.



Electron neutrinos produced from the fusion process inside the Sun are one of the most intensive natural neutrino sources at the Earth, which were first observed at Homestake~\cite{Davis:1968cp}. After that, the solar neutrino detection using the charged-current (CC)~\cite{GALLEX:1992gcp,GNO:2000avz,Abazov:1991rx,SAGE:1999uje,Cleveland:1998nv,SNO:2001kpb} and elastic scattering (ES)~\cite{Kamiokande-II:1989hkh,Super-Kamiokande:2001ljr,Borexino:2007kvk,Borexino:2008fkj,Collaboration:2011nga,BOREXINO:2014pcl,BOREXINO:2020aww} channels have been achieved in various solar neutrino experiments. Meanwhile, the SNO experiment has made the first ever neutral current (NC)~\cite{SNO:2002tuh,SNO:2003bmh,SNO:2008gqy} detection of solar neutrinos and provided the direct test of the standard solar model (SSM)~\cite{Bahcall:1998wm,Bahcall:2000nu,Bahcall:2004fg,Vinyoles:2016djt}.
Inspired by the latest CE$\nu$NS observation, it 
would be encouraging and important to detect the CE$\nu$NS process with solar neutrinos, which is a channel of pure NC detection, and important play ground for the new physics beyond the SM~\cite{Dutta:2019oaj,Billard:2014yka,Cerdeno:2016sfi,Harnik:2012ni}. Moreover, the promising prospect for the CE$\nu$NS detection with solar neutrinos lies in the rapid developments~\cite{Billard:2021uyg,Liu:2017drf,Schumann:2019eaa} of the direct detection of weakly interacting massive particles (WIMPs) as the DM candidate, since the nuclear recoil signals from direction detection of WIMPs and the CE$\nu$NS detection of solar neutrinos are both located at the region from keV to tens of keV, in which high detection efficiency and extremely low background levels have been obtained in current and future DM direct detection programs.  

Thus far, DM direct detection experiments are entering the phase of the multi-ton scale, such as PandaX-4T~\cite{PandaX:2018wtu}, XENON-nT~\cite{XENON:2020kmp}, and LZ~\cite{LZ:2019sgr} and DARWIN~\cite{DARWIN:2016hyl} for the Liquid Xenon (Xe) detectors, and 
DarkSide-20k~\cite{DarkSide-20k:2017zyg} and ARGO~\cite{Billard:2021uyg} for the Liquid Argon (Ar) detectors. Just recently, the PandaX-4T Collaboration has released the first DM search using data of the commissioning run~\cite{PandaX-4T:2021bab}, achieving the currently lowest limit at the DM mass of around 30 GeV. In addition, in the low-mass region of WIMPs, the low-threshold detectors with relatively lighter target nuclei are more advantageous, where 
experiments with cryogenic bolometers are rapidly growing in both the detector size and performance, such as SuperCDMS~\cite{SuperCDMS:2016wui} and EDELWEISS~\cite{EDELWEISS:2017uga} using germanium (Ge) or silicon (Si) as the target. Taking the light mediators of universal scalar, vector and axial types as representatives of new-physics models, in this work, we are going to study the detection potential of coherent elastic solar neutrino nucleus scattering at DM direct detection experiments. Based on the aforementioned experimental plans, and assuming several simplified experimental benchmarks with the Xe, Ar, Ge and Si targets, we present the sensitivity of light mediators as a function of the mediator mass and the coupling strength. Meanwhile we shall also derive the exclusion limits from the recent results of XENON-1T~\cite{XENON:2020gfr}.

The Muon $(g-2)$ Collaboration at Fermi National Laboratory~\cite{Muong-2:2021ojo} has just published a new result on the anomalous muon magnetic moment, and when combined with the old result from Brookhaven National Laboratory~\cite{Muong-2:2006rrc}, induces $4.2\sigma$ inconsistency with the SM prediction~\cite{Aoyama:2020ynm}. Among the numerous and diverse solutions~\cite{Athron:2021iuf,Lindner:2016bgg} to the muon $(g-2)$ anomaly, the light vector mediator with the $L_{\mu}-L_{\tau}$ gauge symmetry is regarded as a viable and simple model~\cite{Cadeddu:2021dqx,Zhou:2021vnf,Ko:2021lpx,Hapitas:2021ilr}. In this regard, we are going to investigate the $L_{\mu}-L_{\tau}$ model using the coherent elastic solar neutrino nucleus scattering in current and future direct detection experiments in this work.


The plan of this work is as follows. In Section II, we present the general framework of the analysis, including the theoretical calculation of the CE$\nu$NS cross section in the presence of light mediators, setups of simplified experimental scenarios, and statistical analysis method. In Section III, the numerical analysis results and discussions are illustrated. Finally we give the concluding remarks in Section IV.


\section{General Framework}

In this part, we present the general framework of the analysis, including the theoretical calculation of the CE$\nu$NS cross section in the presence of light mediators, setups of simplified experimental scenarios, and the statistical analysis method.

\subsection{CE$\nu$NS in the presence of light mediators}
\label{sec:cross}

For the CE$\nu$NS between a neutrino with the energy $E_{\nu}$ and a nucleus with $Z$ protons and $N$ neutrons, the cross section in the SM can be written as~\cite{Drukier:1984vhf,Barranco:2005yy,Patton:2012jr} 
\begin{equation}
\begin{aligned}
\frac{\dif{\sigma_{\mathrm{SM}}}}{\dif{T}}(E_{\nu},T)=&\frac{G_{F}^2  M}{\pi}\left[\left(1-\frac{MT}{2E_{\nu}^2}+\frac{T}{E_{\nu}}\right)(Q^{V}_{\mathrm{SM}})^2\right.\\
&\left.+\left(1+\frac{MT}{2E_{\nu}^2}+\frac{T}{E_{\nu}}\right)(Q^{A}_{\mathrm{SM}})^2-2\left(\frac{T}{E_{\nu}}\right)Q^{V}_{\mathrm{SM}}Q^{A}_{\mathrm{SM}}\right]+\mathcal{O}\left({\frac{T^2}{E^2_{\nu}}}\right)\,,
\end{aligned}
\label{eq:csSM} 
\end{equation}
where $T$ is the kinetic energy of nuclear recoil, $M$ is the nucleus mass, $G_{F}$ is the Fermi constant, and the vector and axial weak charge $Q^{V}_{\mathrm{SM}}$ and $Q^{A}_{\mathrm{SM}}$ are given as
\begin{equation}
Q^{V}_{\mathrm{SM}}=
[g_{V}^{p} Z +g_{V}^{n} N]F_{V}\left(|\vec{q}|^{2}\right)\quad{\rm and}\quad
Q^{A}_{\mathrm{SM}}=
[g_{A}^{p}(Z^{+}-Z^{-}) +g_{A}^{n}(N^{+}-N^{-})]F_{A}\left(|\vec{q}|^{2}\right),
\end{equation}
where $|\vec{q}|^{2}=2MT$, $Z^{\pm}$ and 
$N^{\pm}$ are the numbers of protons and neutrons with spin up (spin down) respectively.
$F_{V}\left(|\vec{q}|^{2}\right)$ and $F_{A}\left(|\vec{q}|^{2}\right)$  are the vector and axial form factors of nucleon distributions in the nucleus respectively. In this work, we neglect the tiny difference between the proton and neutron form factors and employ the proton radii from Ref.~\cite{PhysRevLett.125.141301} with the Helm parameterization~\cite{Helm:1956zz} for all nuclei in the calculation of the next section.
$g_{V}^{p}$ and $g_{V}^{n}$ are the vector neutrino-proton and neutrino-neutron couplings in the SM respectively, which are given as 
\begin{equation}
g_{V}^{p}=-2\sin^{2}\theta_{W}+\frac{1}{2}\simeq0.0229,\quad\quad g_{V}^{n}=-\frac{1}{2}\,,
\end{equation}
where $\theta_{W}$ is the weak mixing angle at low momentum transfer and the radiative corrections have been neglected~\cite{Cadeddu:2020lky}.
Meanwhile, $g_{A}^{p}$ and $g_{A}^{n}$ are respectively the axial neutrino-proton and neutrino-neutron couplings, which can be calculated as 
\begin{equation}
g_{A}^{p}\simeq\sum_{q=u, d, s}g_{A}^{q}\Delta^{p}_{q}=\frac{1}{2}(\Delta^{p}_{u}-\Delta^{p}_{d}-\Delta^{p}_{s}),\quad\quad g_{A}^{n}\simeq\sum_{q=u, d, s}g_{A}^{q}\Delta^{n}_{q}=\frac{1}{2}(\Delta^{n}_{u}-\Delta^{n}_{d}-\Delta^{n}_{s})\,,
\end{equation}
Where $\Delta^{p}_{q}$ and $\Delta^{n}_{q}$ are the axial charges of quarks in the nucleons~\cite{Cirelli:2013ufw}, and only the contributions of three light-flavor quarks are considered.
Note that the axial contribution in the SM can be neglected for most of the nuclei since the ratio of the axial to vector contributions is evaluated to be at the order of $1/(N+Z)$. Thus the SM axial contribution to the CE$ \nu $NS process will not be considered in following calculation. If there are several isotopes for the target nucleus, a weighted average of the cross sections according to their natural abundance will be used.

In order to describe the new physics at a very low energy scale in CE$\nu$NS, we use the effective field theory described in Ref.~\cite{Cirelli:2013ufw} and extend the SM with the flavor-universal scalar (S), vector (V) and axial (A) light mediators, with the extended Lagrangian listed below:
\begin{align}
	\mathcal{L}_{\text {S}}=\phi\left(g_{\phi}^{q S} \bar{q} q+g_{\phi}^{\nu S} \bar{\nu}_{R} \nu_{L}+\text { h.c. }\right),\\
	\mathcal{L}_{\text {V}}=Z_{\mu}^{\prime}\left(g_{Z^{\prime}}^{q V} \bar{q} \gamma^{\mu} q+g_{Z^{\prime}}^{\nu V} \bar{\nu}_{L} \gamma^{\mu} \nu_{L}\right), \\
	\mathcal{L}_{\text {A}}=Z_{\mu}^{\prime}\left(g_{Z^{\prime}}^{q A} \bar{q} \gamma^{\mu} \gamma^{5}q+g_{Z^{\prime}}^{\nu A} \bar{\nu}_{L} \gamma^{\mu} \gamma^{5} \nu_{L}\right),
\end{align}
where $g_{\phi}^{f S}$ is the scalar coupling to the fermion $f=(u,d,\nu)$
of the scalar mediator $\phi$ with the mass $M_{\phi}$, while 
$g_{Z^{\prime}}^{f V}$ ($g_{Z^{\prime}}^{f A}$) is the (axial) vector coupling to the fermion of the vector mediator $Z^{\prime}$ with the mass $M_{Z^{\prime}}$.
Note that flavor-universal couplings to the up and down quarks (i.e., $g_{Z^{\prime}}^{u V}=g_{Z^{\prime}}^{d V}=g_{Z^{\prime}}^{q V}$ and $g_{Z^{\prime}}^{u A}=g_{Z^{\prime}}^{d A}=g_{Z^{\prime}}^{q A}$) have been assumed in all the considered scenarios. 

In the presence of new light mediators, the SM cross section will be modified.
To begin with, the scalar mediator contributes an incoherent cross-section term in addition to the SM cross section as: 
\begin{equation}\label{cs_total_scalar}
	\dfrac{d\sigma_{\mathrm{SM+S}}}{d T}(E_{\nu},T)
	=
	\dfrac{d\sigma_{\mathrm{SM}}}{d T_\mathrm{}}(E_{\nu},T)
	+
	\dfrac{d\sigma_{\mathrm{S}}}{d T_\mathrm{}}(E_{\nu},T),
\end{equation} 
where the scalar contribution is derived as
\begin{equation}
\dfrac{d\sigma_{\mathrm{S}}}{d T_\mathrm{}}(E_{\nu},T)
	=\frac{M^{2}}{4 \pi}
		\frac{T_{}}{E_{\nu}^{2}}
	 \frac{({Q}^{S}_{\phi})^{2}}{\left(|\vec{q}|^{2}+M_{\phi}^{2}\right)^{2}},
	 \label{cs_scalar}
\end{equation}
with the scalar charge given as~\cite{AristizabalSierra:2019ykk}
\begin{equation}\label{Q_scalar}
	{Q}^{S}_{\phi}=\left[Z\sum_{q=u, d, s}  \frac{m_{p}}{m_{q}} f_{T_{q}}^{p}+N  \sum_{q=u, d, s} \frac{m_{n}}{m_{q}} f_{T_{q}}^{n}\right]g_{\phi}^{\nu S}g^{q S}_{\phi}, 
\end{equation}
where $f_{T_{q}}^{p,n}$ are the hadronic form factors, and obtained from the chiral perturbation theory~\cite{Ellis:2018dmb,Hoferichter:2015dsa}.

On the other hand, the light vector mediator will contribute to the CE$\nu$NS cross section in a coherent way, with a direct modification to the vector weak charge as
\begin{equation}\label{cs_total_vector}
\begin{aligned}
\dfrac{d\sigma_{\mathrm{SM+V}}}{d T}(E_{\nu},T)
=&\frac{G_{F}^2 M}{\pi}\left[\left(1-\frac{MT}{2E_{\nu}^2}+\frac{T}{E_{\nu}}\right)(Q^{V}_{\mathrm{SM+V}})^2\right.\\
&\left.+\left(1+\frac{MT}{2E_{\nu}^2}+\frac{T}{E_{\nu}}\right)(Q^{A}_{\mathrm{SM}})^2-2\left(\frac{T}{E_{\nu}}\right)Q^{V}_{\mathrm{SM+V}}Q^{A}_{\mathrm{SM}}\right]+\mathcal{O}\left({\frac{T^2}{E^2_{\nu}}}\right)\,,
\end{aligned}
\end{equation} 
with 
\begin{equation}\label{Q_SMV}
	Q^{V}_{\mathrm{SM+V}}=
	\left[Q^{V}_{\mathrm{SM}} - 
	\frac{3g^{\nu V}_{Z^{\prime}} g^{q V}_{Z^{\prime}}(Z + N)}
	{\sqrt{2} G_{F}\left(|\vec{q}|^{2}+M_{Z^{\prime}}^{2}\right)} 
	\right]\,. 
\end{equation}

Furthermore, it is also the coherent contribution for the light axial mediator, but with the SM axial vector part:
\begin{equation}\label{cs_total_axial}
\begin{aligned}
\dfrac{d\sigma_{\mathrm{SM+A}}}{d T}(E_{\nu},T)
=&\frac{G_{F}^2 M}{\pi}\left[\left(1-\frac{MT}{2E_{\nu}^2}+\frac{T}{E_{\nu}}\right)(Q^{V}_{\mathrm{SM}})^2\right.\\
&\left.+\left(1+\frac{MT}{2E_{\nu}^2}+\frac{T}{E_{\nu}}\right)(Q^{A}_{\mathrm{SM+A}})^2-2\left(\frac{T}{E_{\nu}}\right)Q^{V}_{\mathrm{SM}}Q^{A}_{\mathrm{SM+A}}\right]+\mathcal{O}\left({\frac{T^2}{E^2_{\nu}}}\right)\,,
\end{aligned}
\end{equation} 
with 
\begin{equation}\label{Q_SMA}
	Q^{V}_{\mathrm{SM+A}}=
	\left\{Q^{A}_{\mathrm{SM}} + 
	\frac{g^{\nu A}_{Z^{\prime}} g^{q A}_{Z^{\prime}}\left[(\sum_{q}\Delta^{(p)}_{q})(Z^{+}-Z^{-}) + (\sum_{q}\Delta^{(n)}_{q}) (N^{+}-N^{-})\right]}
	{\sqrt{2} G_{F}\left(|\vec{q}|^{2}+M_{Z^{\prime}}^{2}\right)} 
	\right\}\,. 
\end{equation}
Note that $\sum_{q}\Delta^{(p)}_{q}=\sum_{q}\Delta^{(n)}_{q}\simeq0.3$~\cite{Cerdeno:2016sfi}, thus the additional axial charge will be directly related to the nuclear spin.

Finally let us consider the flavor-dependent $L_{\mu}-L_{\tau}$ model of light vector mediators.
Since $Z^{\prime}$ only interacts with the muonic or tauonic leptons, but not directly with quarks, there is no tree-level contributions to the CE$\nu$NS cross section, but loop-level contributions exist with the virtual $\mu$ and $\tau$ exchange through kinetic mixing involving photons, where the vector weak charge in Eq.~(\ref{cs_total_vector}) is altered as
\begin{equation}\label{Q_SMmutau}
	Q^{V}_{\mathrm{\mu\tau}}=
	\left[Q^{V}_{\mathrm{SM}} - \frac{\alpha (g^{\mu\tau}_{Z^{\prime}})^2 }{3 \sqrt{2} \pi G_{F}}\log{\frac{m_{\tau}^{2}}{m_{\mu}^{2}}\frac{Z}{\left|\vec{q} \right|^{2}+M_{Z^{\prime}}^{2} }} 
	\right]\,, 
\end{equation}
where $\alpha$ is the fine structure constant of electromagnetic interactions, $m_{\mu}$ and $m_{\tau}$ are masses of $\mu$ and $\tau$ respectively.
Note that only the proton part of the vector weak charge is modified because of the presence of the photon in the loop, while the contributions of neutrons remain unchanged.

\begin{figure}
	\centering
		\includegraphics[scale=0.25]{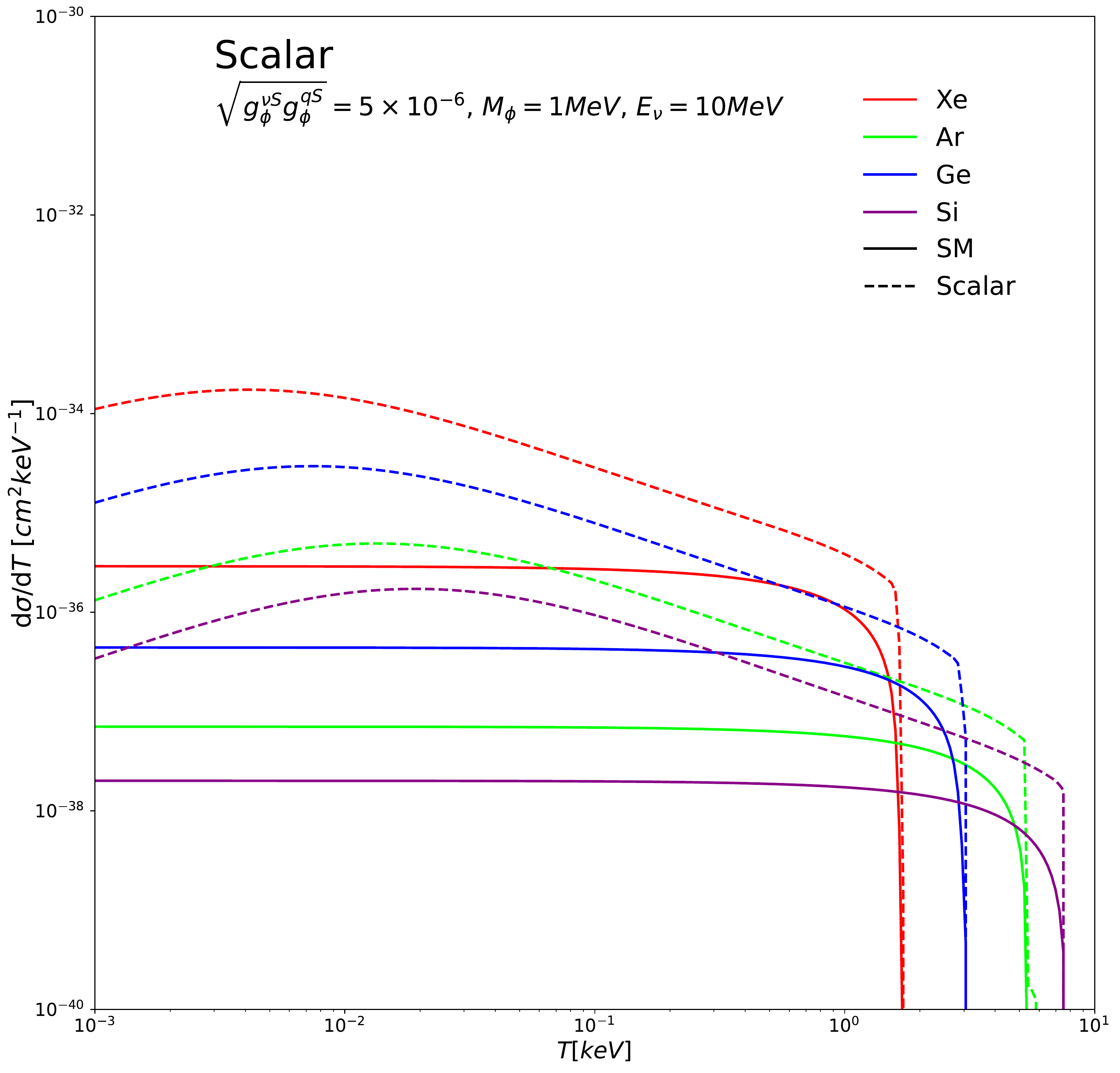}
		\includegraphics[scale=0.25]{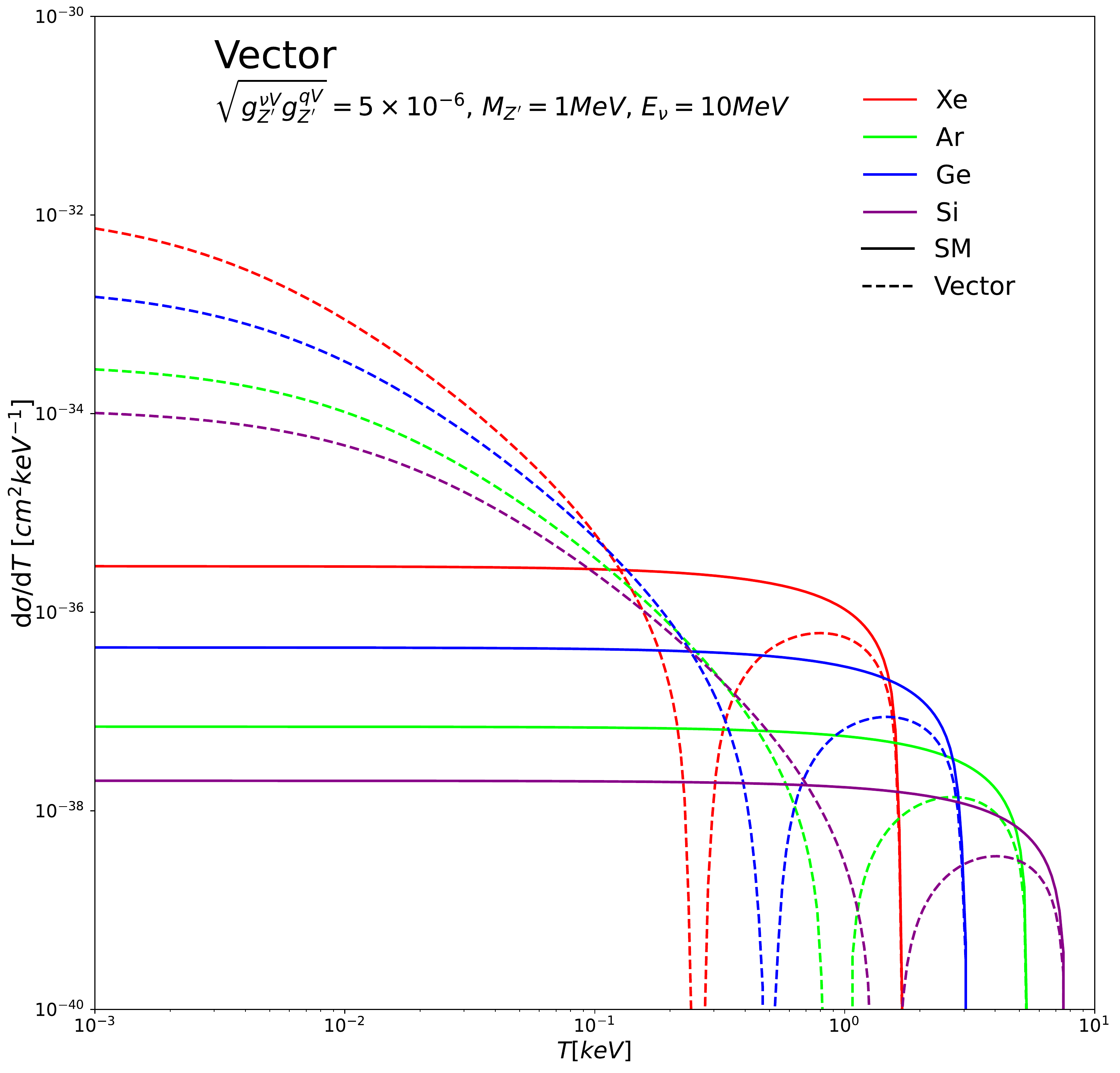}
		\includegraphics[scale=0.25]{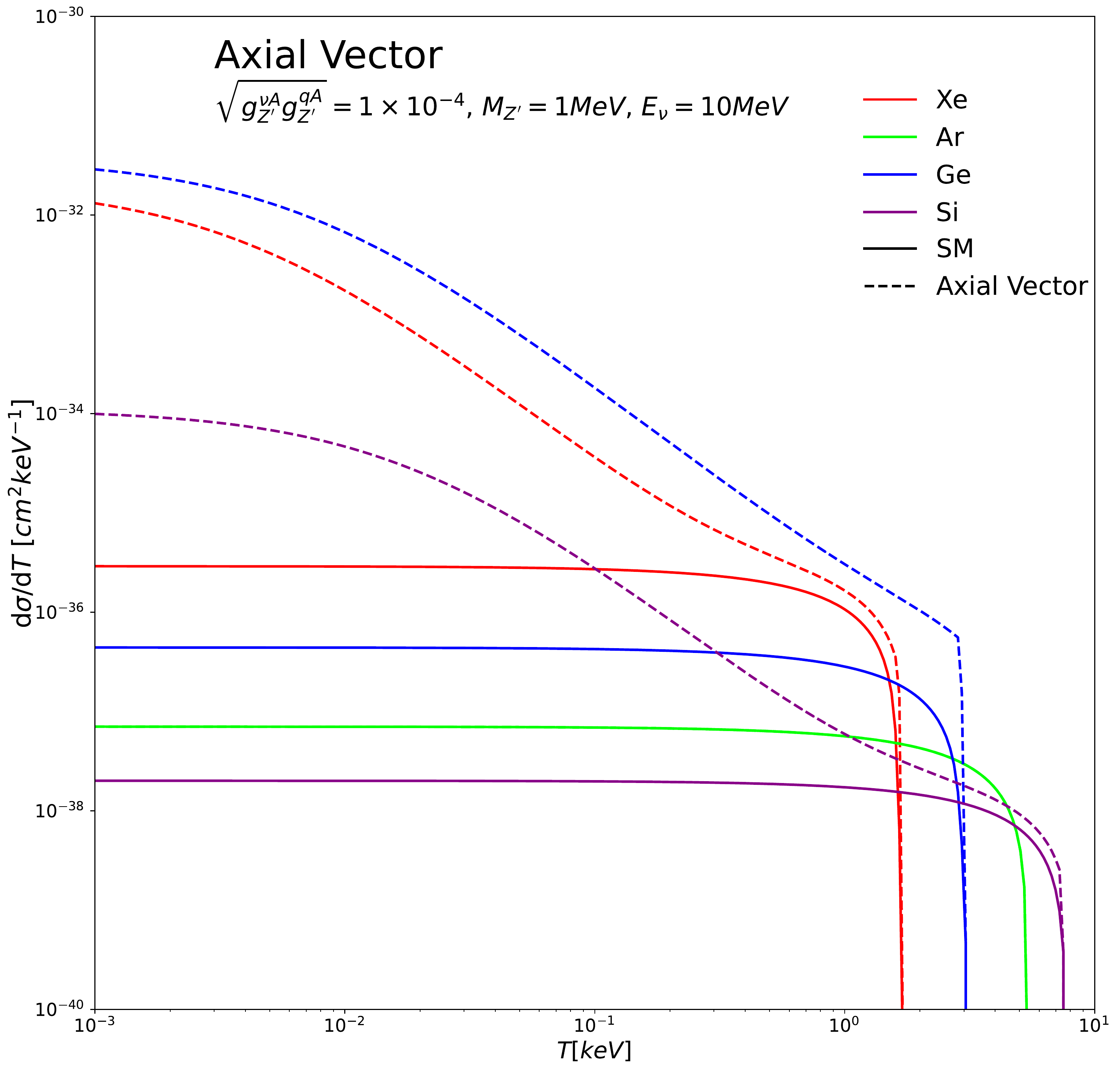}
		\includegraphics[scale=0.25]{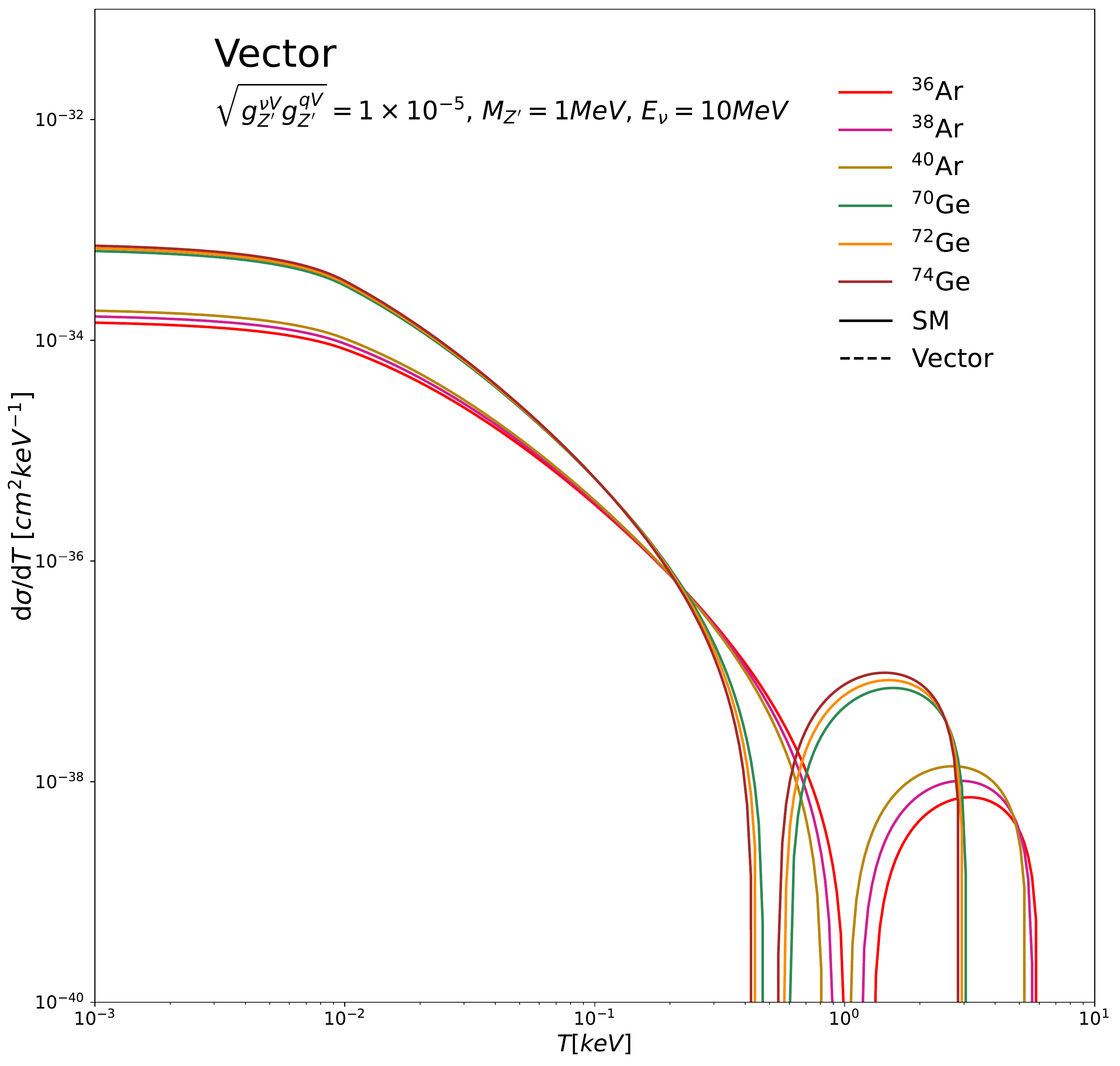}
		\caption{The cross section as a function of the nuclear recoil energy for the targets of Si, Ar, Ge and Xe with different light mediators (top left: scalar; top right: vector; bottom left: axial vector), in which the weighted average has been performed according to the natural abundance of target isotopes. In the bottom right panel, the cross sections as a function of the nuclear recoil energy for different isotopes of Ar and Ge are illustrated. The neutrino energy is set to 10 MeV and the mediator mass is set to 1 MeV for all the calculations. The values of interaction couplings have been specified in each plot.}
		\label{fig:cross-setion}
\end{figure}

Before finishing this part, we want to illustrate properties of the CE$\nu$NS cross section in the presence of light mediators, some of which are also discussed, for example, in Ref.\cite{Boehm:2018sux,Bertuzzo:2017tuf}. In Fig.~\ref{fig:cross-setion}, the cross sections are shown as a function of the nuclear recoil energy for the targets of silicon (Si), argon (Ar), germanium (Ge) and xenon (Xe) with different light mediators (top left: scalar; top right: vector; bottom left: axial vector), in which the weighted average has been performed according to the natural abundance of isotopes of the target. In the bottom right panel, the cross sections as a function of the nuclear recoil energy for different isotopes of Ar and Ge are illustrated. The neutrino energy is set to 10 MeV and the mediator mass is set to 1 MeV for all the calculations. The values of the interaction couplings have been specified in each plot.
From the figure, we can observe that the scalar mediator always enhances the CE$\nu$NS cross section, but there are strong cancellation regions for the vector mediator. The reason is that scalar mediator contributes an incoherent component of the cross section, but vector mediator may significantly decrease the vector weak charge because of the cancellation. From the bottom right panel, one can note that the cancellation depends on the types of nuclear target, as well as the different isotopes. For the axial vector mediator, its contribution is also additive since the SM cross section is vector dominant due to coherent enhancement of heavy nuclei.
Note that the contribution for Ar is vanishing because of the zero total spin.

\subsection{Experimental Scenarios}

Now we are going to introduce the setup of experimental scenarios considered in this work. After investigating the current and future generations of DM direct detection experiments with various target materials and detector technologies, we summarize the following observations:

\begin{itemize}

\item Firstly, Xe-based experiments have achieved compelling DM results for the WIMPs mass above tens of GeV~\cite{XENON100:2010cgk,LUX:2013afz,PandaX-II:2016vec}. Currently, we have the new generation of Xe-based DM direct detection experiments, including PandaX-4T~\cite{PandaX:2018wtu}, XENON-nT~\cite{XENON:2020kmp}, and LZ~\cite{LZ:2019sgr}, reaching the multiple ton scale, and with promising prospect to first detect the CE$\nu$NS process with solar neutrinos. In the future, the flagship experiment DARWIN~\cite{DARWIN:2016hyl} is planed to deploy 50 tons of xenon as an ultimate experiment for the WIMPs search. 

\item Secondly, Ar-based experiments have the advantage of high recoil energies because of lighter nucleus mass, which have achieved considerable DM results~\cite{BKar,DEAP3600}. An excellent representive of new generation Ar-based experiments is Darkside-20k~\cite{DarkSide-20k:2017zyg}, which is planed to deploy 40 ton LAr for the DM and solar neutrino detection. In the far future, there is an idea of ARGO~\cite{Billard:2021uyg}, which will increase the mass of Ar to 400 tons.  

\item Thirdly, there are also plenty of low threshold DM detectors~\cite{CDMS:2013juh,SuperCDMS:2014cds,Aramaki:2016spe,CDEX:2018lau,DAMIC:2021esz,SENSEI:2020dpa}, which are designed for the low mass region of WIMPs. Typical examples of the next generation experiments are Super CDMS~\cite{SuperCDMS:2016wui}, EDELWEISS-III~\cite{EDELWEISS:2017uga}, SENSEI~\cite{Tiffenberg:2017aac}.
which have extremely low energy threshold, providing excellent opportunity to constrain light new physics.

\end{itemize}

\begin{table}
	\centering
	\begin{tabular}{ccccc}
		\toprule
		Type& Target & Exposure&  Optimal/Nominal Threshold& Background  \\
		& & [t$\times$year] & [keV] &[t$^{-1}$year$^{-1}$keV$^{-1}$]\\
		\midrule 
		Ge-Gen-II&Ge&0.2&0.04/0.1 &1 \\
		Ge-Future&Ge&2&0.04/0.1&1 \\
		Si-Gen-II&Si&0.2 &0.04/0.1 &1 \\
		Si-Future&Si&2 &0.04/0.1 &1 \\
		Xe-Gen-II&Xe&20 &1/3.5&2 \\
		Xe-Future&Xe&200 &1/3.5&2 \\
		Ar-Gen-II&Ar&200 &1/3.5&2\\
		Ar-Future&Ar&3000 &1/3.5&2 \\
		\bottomrule
	\end{tabular}
	\caption{Experimental scenarios and their typical parameters employed in this work.}
	\label{table:exp}
\end{table}
In this work, motivated by the above investigation, we shall consider the experimental scenarios listed in Table~\ref{table:exp} with four target materials and two levels of target masses, where Gen-II indicates the experiments in the coming years and Future represents those in the far future with much higher target masses. 
For each scenario, we consider a nominal and an optimistic energy threshold in terms of the nuclear recoils. All the experiments are expected to reach 100$\%$ efficiency above the threshold.
Since a detailed background budget is too complicated for the general analysis of new physics, we also simplify as a flat background level based on the consideration in Refs.~\cite{SuperCDMS:2016wui,BKar,BKxe}, which are also listed in Table~\ref{table:exp}.

\subsection{Statistical Method}

In this work, we are interested the CE$\nu$NS interactions of solar neutrinos in the DM direct detection experiments. In general, the event numbers of the solar neutrino CE$\nu$NS process in a certain range of the nuclear recoil energy can be written as
\begin{equation}
N_{i} = \frac{\epsilon}{M} \int_{T_{{i,\rm min}}}^{T_{{i,\rm max}}} \dif{T} \int_{E_{{\rm min}}}^{E_{{\rm max}}} \dif{E_{\nu}} \cdot  \Phi (E_{\nu}) \frac{\dif{\sigma}}{\dif{T}}\,,
\end{equation}
where $\epsilon$ is the exposure of the considered experiment and $M$ is the mass of target nucleus, depending on the type of the experiment. $\Phi(E_{\nu})$ is the solar neutrino fluxes from the standard solar model (SSM)~\cite{Vinyoles:2016djt}. $T$ is the recoil energy, $E_{\nu}$ is the neutrino energy, with $E_{{\rm max}}$ being the maximal neutrino energy and $E_{{\rm min}}$ the minimal neutrino energy for a certain recoil energy which can be written as
\begin{equation}
E_{{\rm min}}=\frac{T}{2}\left(1+\sqrt{1+2\frac{M}{T}}\right)\,.
\end{equation}

To explore the constraints on light mediator models from the solar neutrino CE$\nu$NS process with the considered experimental scenarios, we employ the standard least squares method with the asimov data set for each experimental scenario listed in Table~\ref{table:exp}:
\begin{equation}
    \chi^2=\sum_{i}\frac{[N^{{\rm exp}}_{i}-N^{{\rm pred}}_{i}(\mathbf{p})]^2}{N^{{\rm bkg}}_{i}+N^{{\rm exp}}_{i}}\,,
\end{equation}
where $N^{{\rm exp}}_{i}$ and $N^{{\rm pred}}_{i}(\mathbf{p})$ are the experimental and predicted event numbers from the considered experiment in the $i$th bin and $N^{{\rm bkg}}_{i}$ is the corresponding background. $\mathbf{p}$ is the vector of the physical parameters considered in each model. Note that we have neglect possible systematic uncertainties, which may worsen the results, but the orders of magnitude of the sensitivity should be viable and meaningful.

In this work, we shall discuss the constraints from $^8$B solar neutrino results of XENON-1T~\cite{Xenon}, which presented the 90$ \% $ confidence level (C.L.) upper limit on the $^8$B solar neutrino flux as $\Phi_{90\%}=1.4\times 10^{7} $ cm$^{-2}$s$^{-1}$,
Since the $^8$B solar neutrino flux from the SSM is $ \Phi_{{\rm SSM}}=(5.25 \pm 0.20)\times 10^{6} $ cm$^{-2}$s$^{-1}$, the constraints to any new-physics (NP) effects can be expressed as
\begin{equation}
\frac{\langle N_{{\rm NP}}\left(\mathbf{p}\right) \rangle }{\left\langle N_{{\rm SM}}\right\rangle }  \lesssim \frac{\Phi_{90\%}}{\Phi_{{\rm SSM}}}\,,
\end{equation}
where the average sign denotes the isotopic average performed based on the natural abundance of Xe, and $N_{{\rm NP}}$ and $N_{{\rm SM}}$ are the expected event rates from the NP model and the SM respectively
\begin{equation}
N_{{\rm NP,\;SM}}= \frac{\epsilon}{M} \int_{T_{\rm min}}^{T_{\rm max}} \dif{T} \int_{E_{\rm min}}^{E_{\rm max}} \dif{E_{\nu}} \cdot  \Phi (E_{\nu}) \eta\left(E_{\nu} \right)\frac{\dif{\sigma}_{{\rm NP,SM}}}{\dif{T}}\,.
\end{equation}
The differential cross section for NP models and the SM are already illustrated in section 2.1 and the detector efficiency $\eta\left(E_{\nu} \right)$ is taken from Ref.\cite{Xenon}.

\section{Numerical Results}

In this section, we are going to present numerical analysis results. First we show 
the expected event spectra for each experimental scenario as functions of the recoil energy and energy threshold of the target nuclei.
Then we illustrate the sensitivity of the solar neutrino CE$\nu$NS detection on the flavor-universal scalar, vector and axial vector mediators, as well as the flavor-specified $L_{\mu}-L_{\tau}$ model towards the solution to the $ (g-2)_{\mu}$ anomaly.

\subsection{Expected Event Spectra}

\begin{figure}
	\centering
	\includegraphics[scale=0.38]{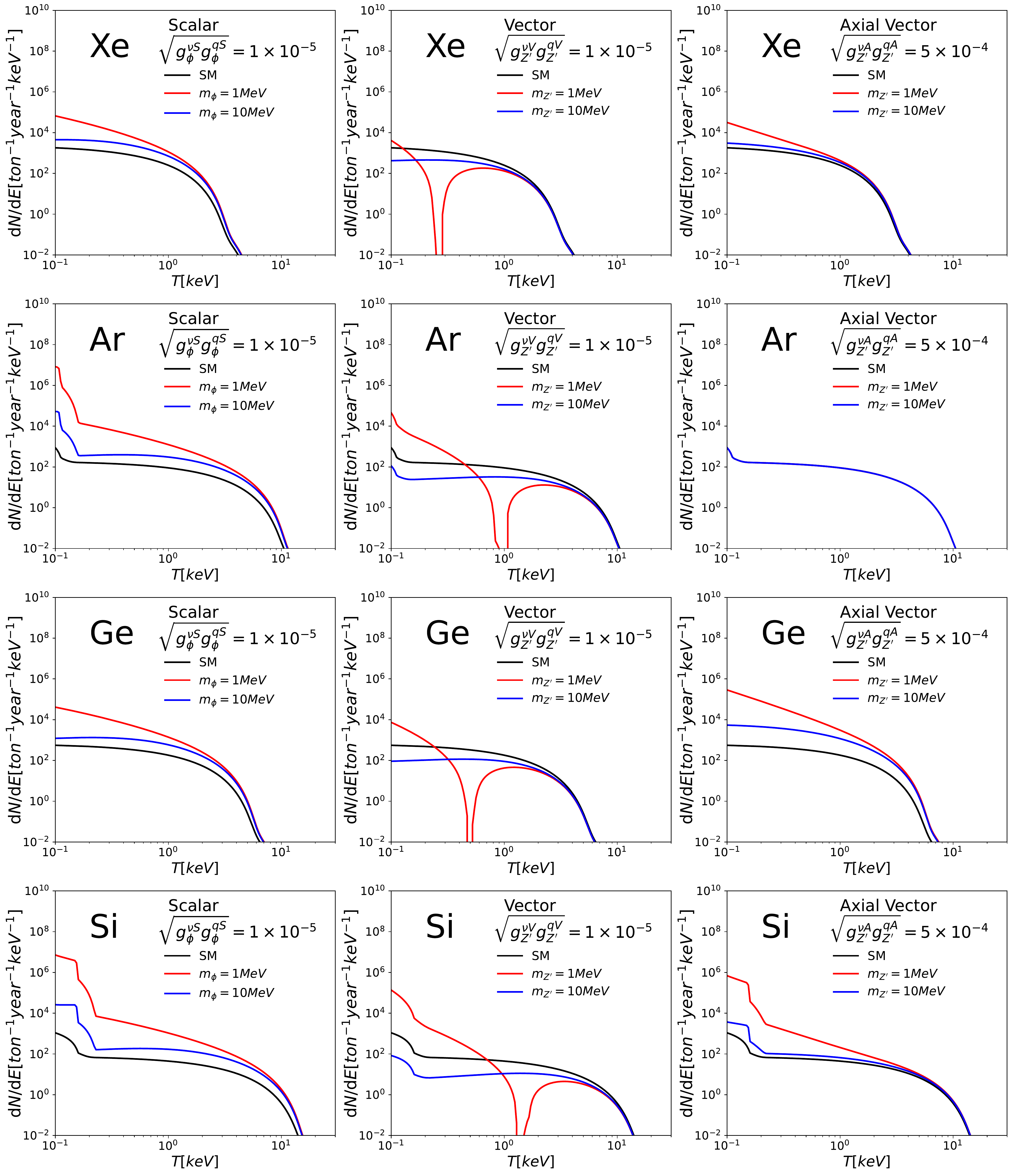}
	\caption{Expected event energy spectra as a function of the nuclear recoil energy for different detector materials and different light mediator models. 
	A weighted average have been performed according to the natural abundance of isotopes in detector material. From top to bottom rows results are shown for Xe, Ar, Ge and Si detectors respectively. From left to right results are shown for the scalar, vector and axial vector mediators respectively. The masses and coupling strength of the light mediators have been specified in each plot.}
	\label{fig:specrecoil}	
\end{figure}
\begin{figure}
	\centering
	\includegraphics[scale=0.38]{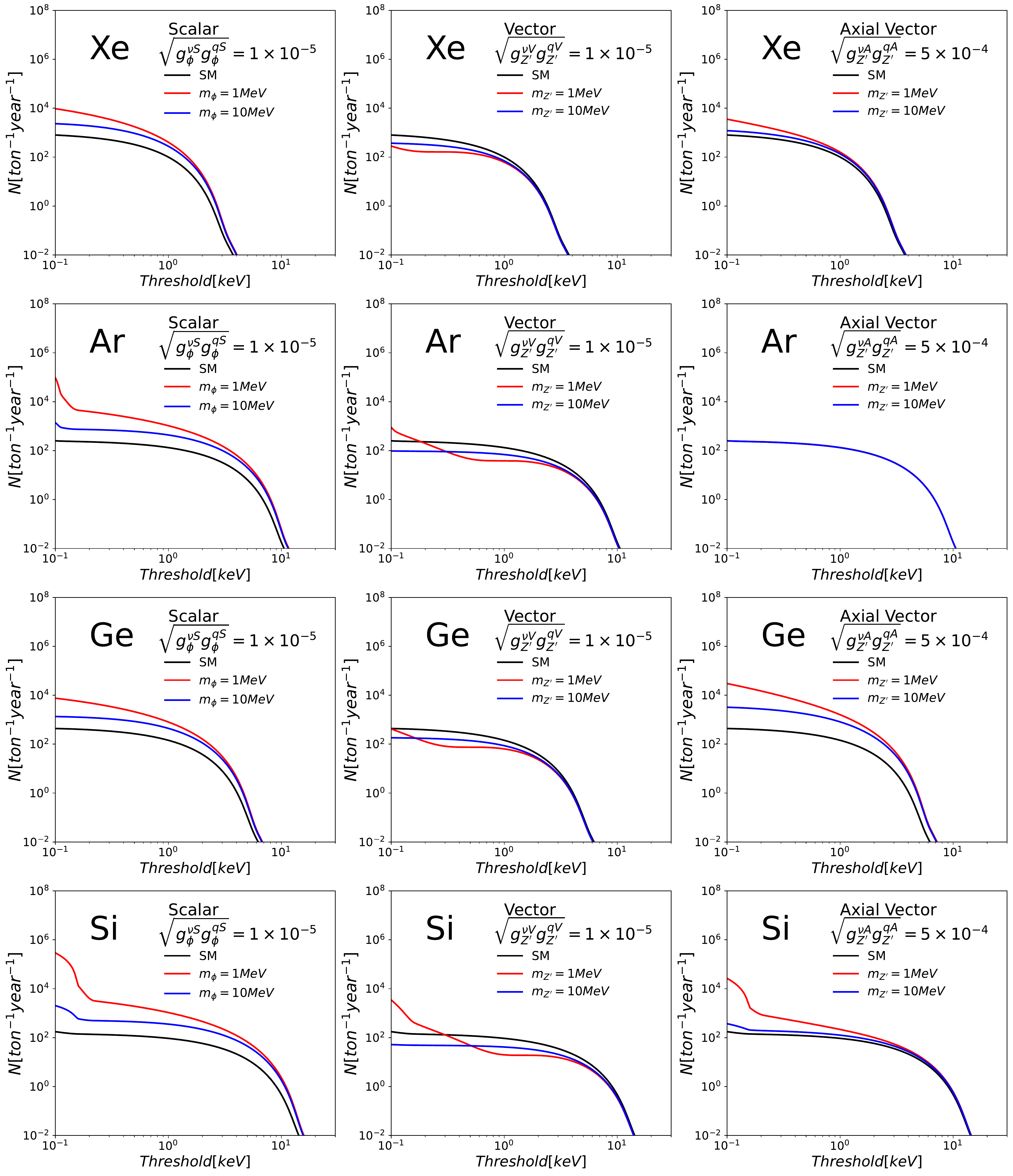}
	\caption{
	Expected event energy spectra as a function of the energy threshold for different detector materials and different light mediator models. A weighted average have been performed according to the natural abundance of isotopes in detector material. From top to bottom rows results are shown for Xe, Ar, Ge and Si detectors respectively. From left to right results are shown for the scalar, vector and axial vector mediators respectively. The masses and coupling strength of the light mediators have been specified in each plot.}
	\label{fig:specthreshold}	
\end{figure}

In Fig.~\ref{fig:specrecoil} and Fig.~\ref{fig:specthreshold}, we illustrate the effects of scalar, vector and axial vector mediators on the expected event energy spectra of the solar neutrino CE$\nu$NS process as functions of the nuclear recoil energy and energy threshold, respectively. A weighted average have been performed according to the natural abundance of isotopes in detector material. From top to bottom rows results are shown for Xe, Ar, Ge and Si detectors respectively. From left to right results are shown for the scalar, vector and axial vector mediators respectively. The masses and coupling strength of the light mediators have been specified in each plot.
The mediator effects would become significant when the recoil energy decreases, which requires detectors with very low threshold for an effective observation. 
Since the maximum nuclear recoil energy for a certain neutrino energy is given by
\begin{equation}
T_{\rm max}=\frac{2E_{\nu}^{2}}{2E_{\nu}+M}\,,
\end{equation}   
therefore lighter nuclear target will result in a higher maximum recoil energy and a relative higher threshold can be acceptable. By carefully looking into the properties of the figures, several comments are provided as follows.

\begin{itemize}

\item For the scalar interaction, the NP effect becomes significant when the recoil energy reaches 0.1 keV and increases steadily as recoil energy decreases. This kind of interaction dramatically enhances the event rate of the CE$\nu$NS process at a recoil energy of $\mathcal{O}(1-10)$ eV while shows almost no effect when recoil energy is higher than 1 keV because of a $T^{-1}$ factor in the cross section, which makes it an effective and universal way to improves measurements of the low energy events.

\item For the vector interaction, the NP and SM contributions may lead to cancellation at a certain recoil energy and create a steep valley in the spectra. For detectors containing several isotopes like those considered in this work, the location and bottom value of cancellation should be averaged based on the weights and spectra of different isotopes as shown in Fig.~\ref{fig:cross-setion}. Since the cancellation is sensitive to the coupling strength, it could be an effective approach to constrain the coupling by using the cancellation location.

\item For the axial vector interaction, the NP contribution is related to the nuclear spin and for some detector materials with no isotopes with nonzero nuclear spin, the NP effect will be vanishing. For example, as shown in Fig.~\ref{fig:cross-setion}, the axial vector interaction cannot be observed in Ar detectors since all the long-life isotopes of Ar are with zero spin. The axial vector enhancement to CE$\nu$NS is generally suppressed by a $T^{-1}$ factor in the cross section like the scalar interaction while it is also related to the abundance and nuclear spin of isotopes. 

\item The event rate spectra as a function of the energy threshold with three interactions behave generally similar and highlights the significance of low threshold detectors since all three NP interactions are suppressed by the recoil energy and become observable only below 1 keV. For the vector interaction, event rate is slightly lower at certain energy range at $\mathcal{O}(0.1-1)$ keV due to the cancellation. The enhancement of the event rates by scalar and axial vector interactions is similar, while the effect of the axial vector interaction can be strengthened (Ge), weakened (Xe and Si) or canceled (Ar) based on the abundance and nuclear spin of isotopes.

\end{itemize}


\subsection{Constraints on the Light Mediators}
\label{sec:constraint-general}

\begin{figure}
	\centering
	\includegraphics[scale=0.32]{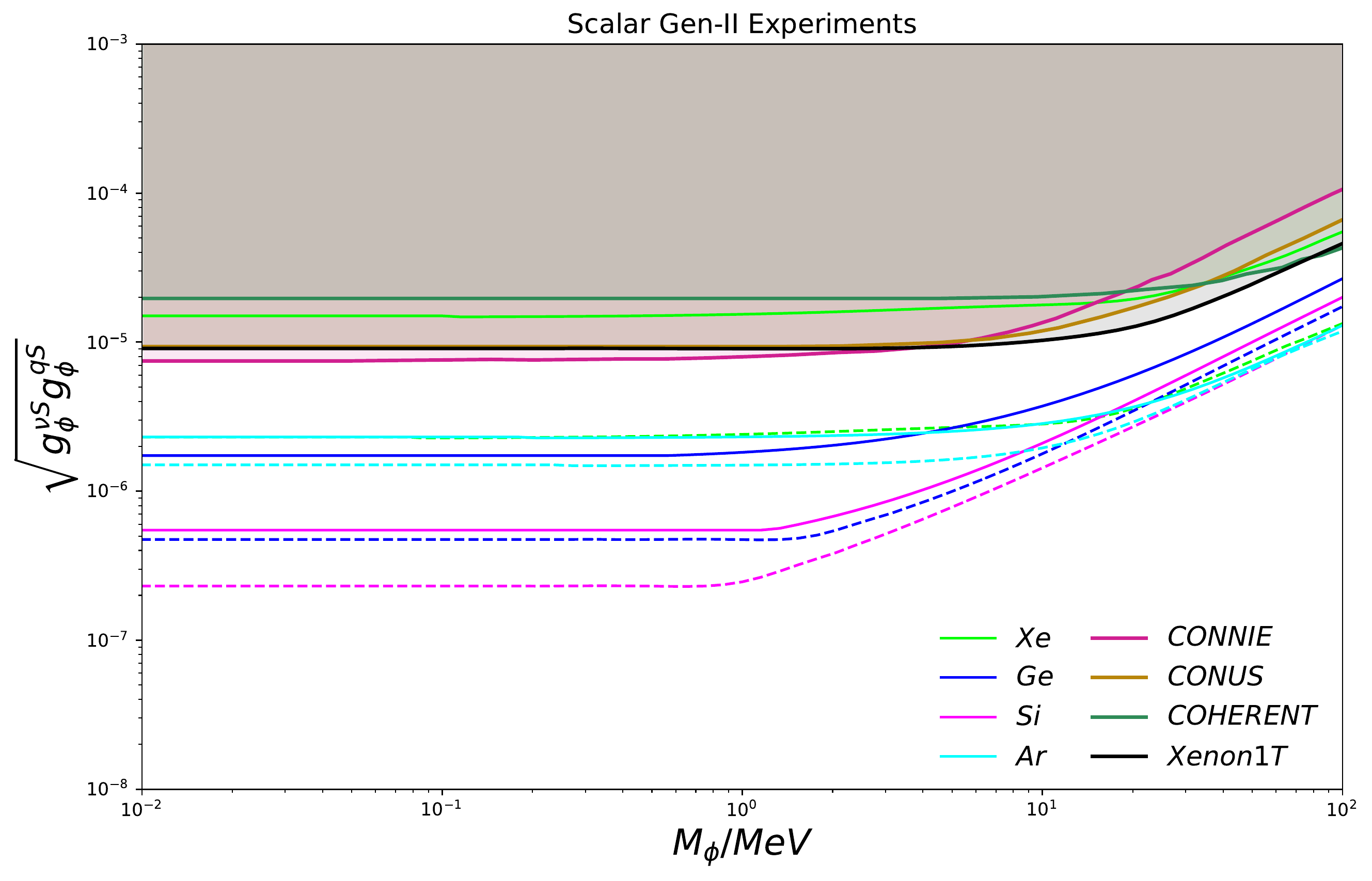}
	\includegraphics[scale=0.32]{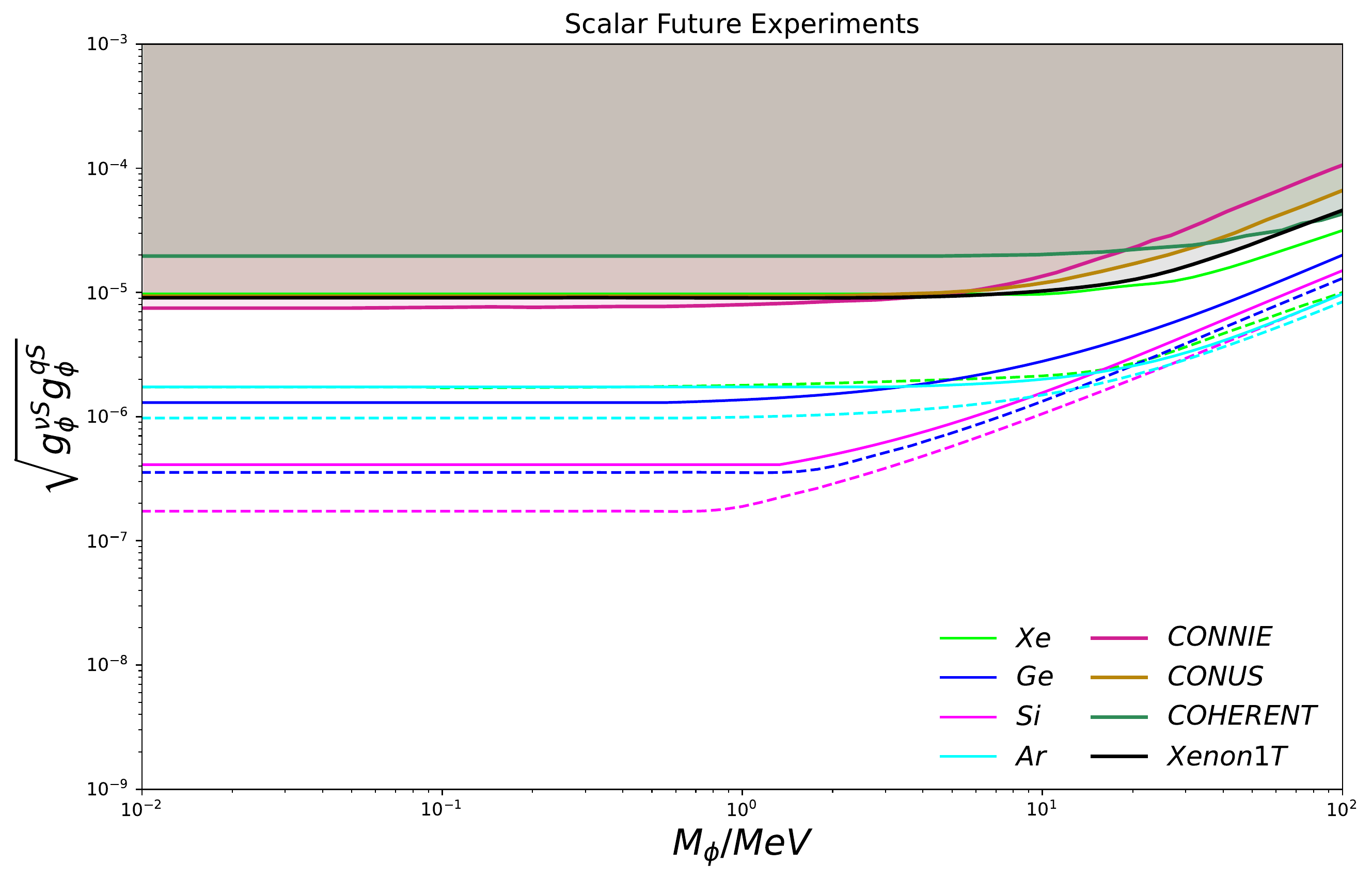}
	\includegraphics[scale=0.32]{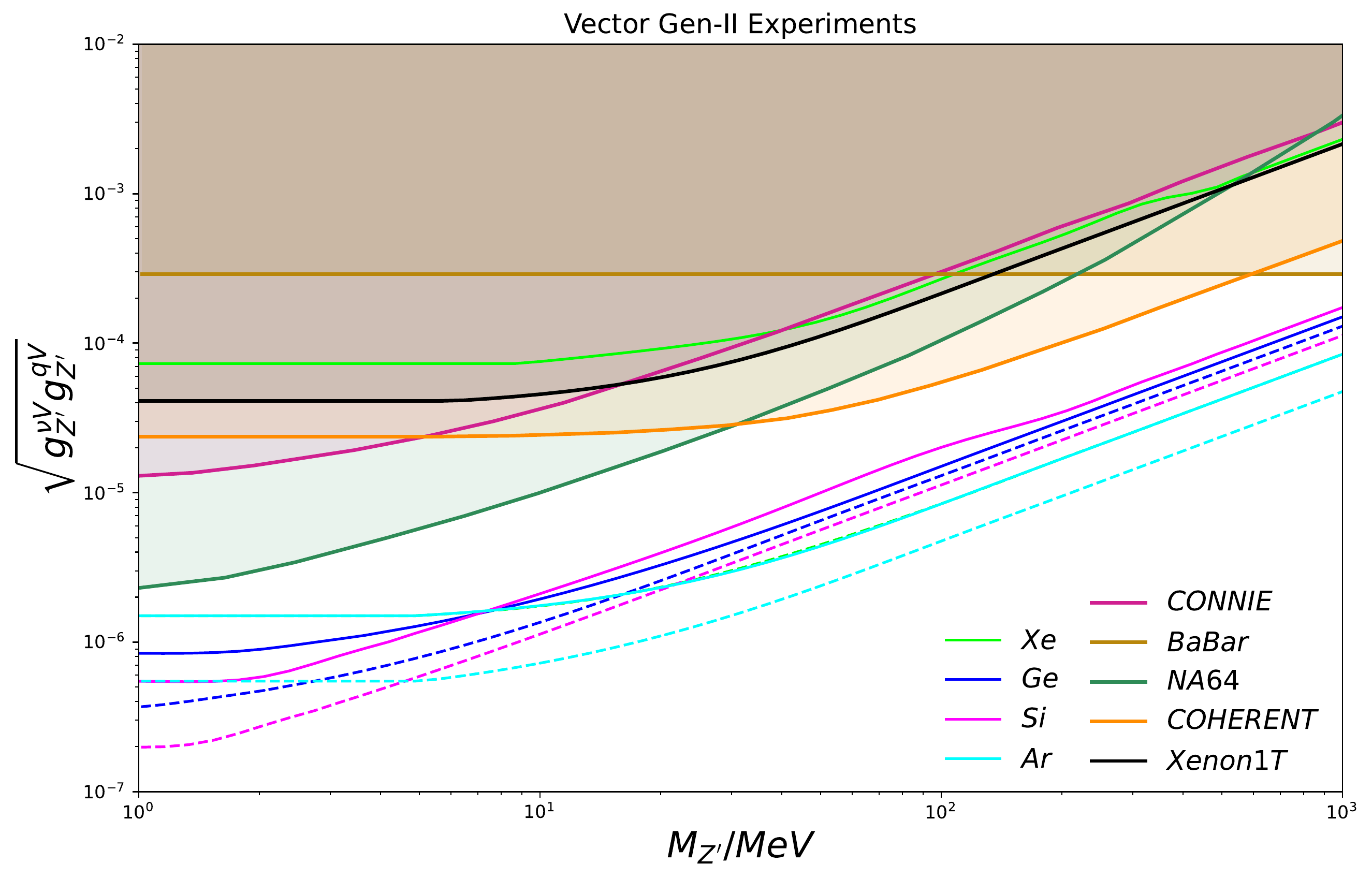}
	\includegraphics[scale=0.32]{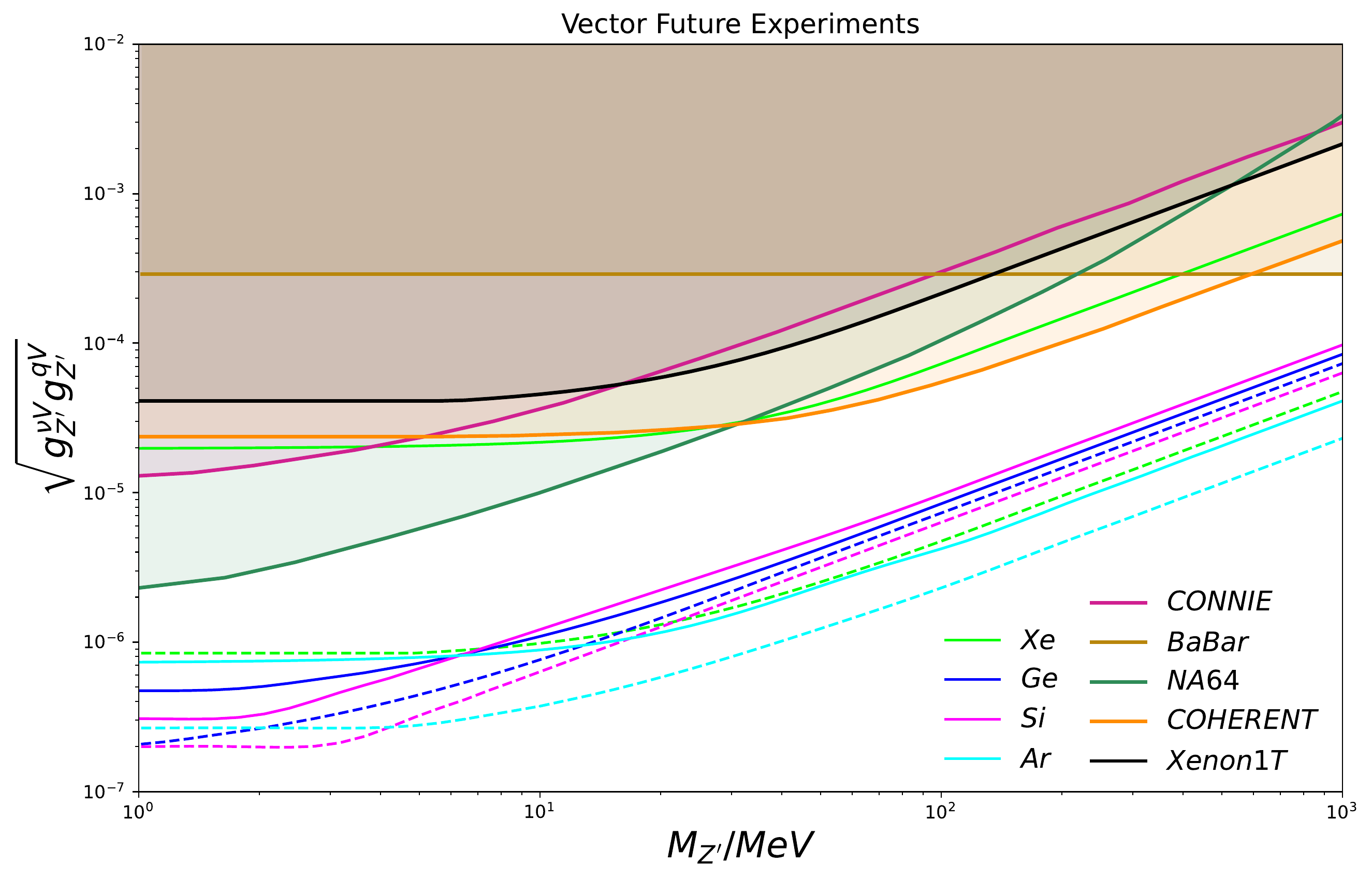}
	\includegraphics[scale=0.32]{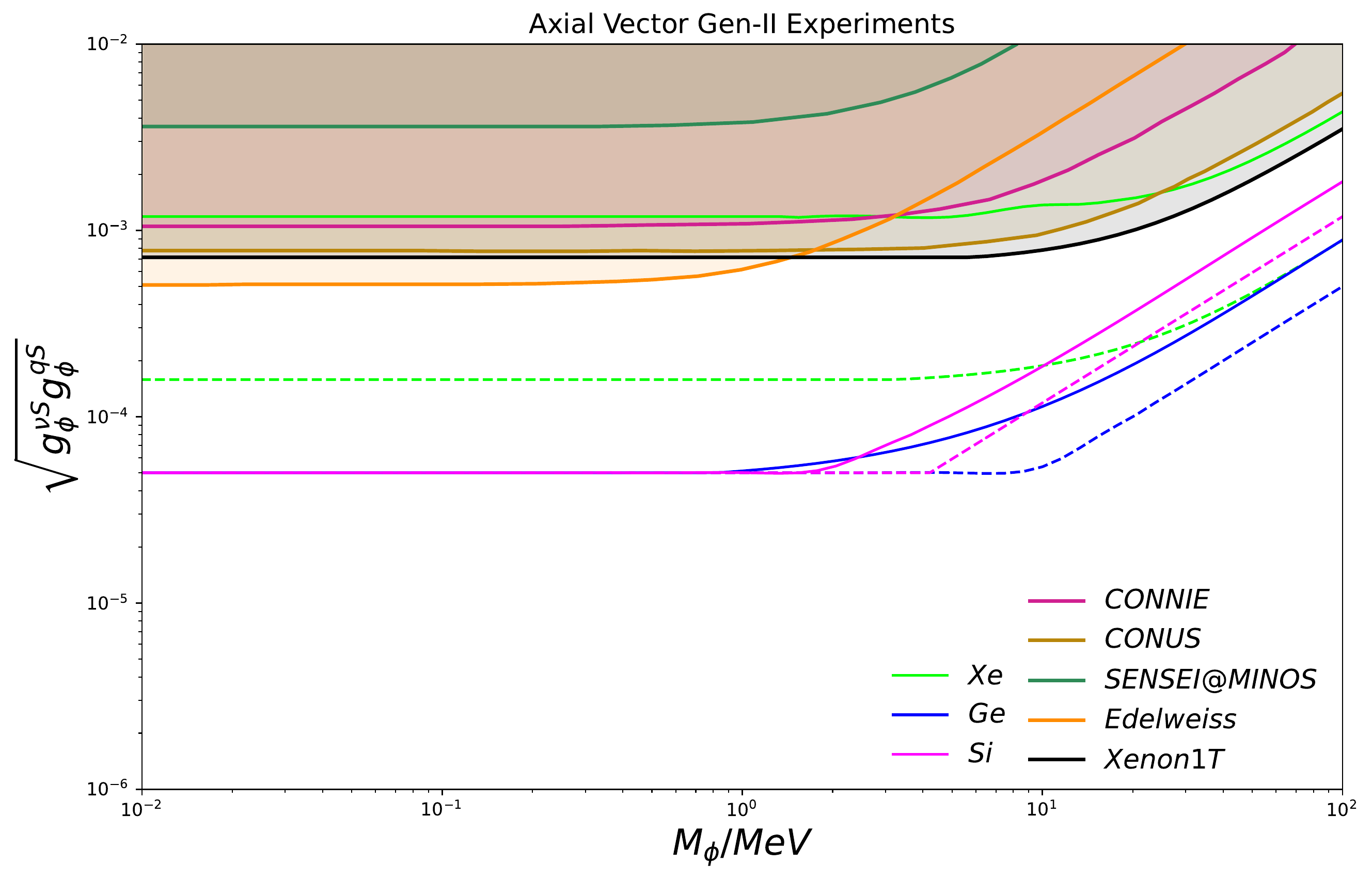}
	\includegraphics[scale=0.32]{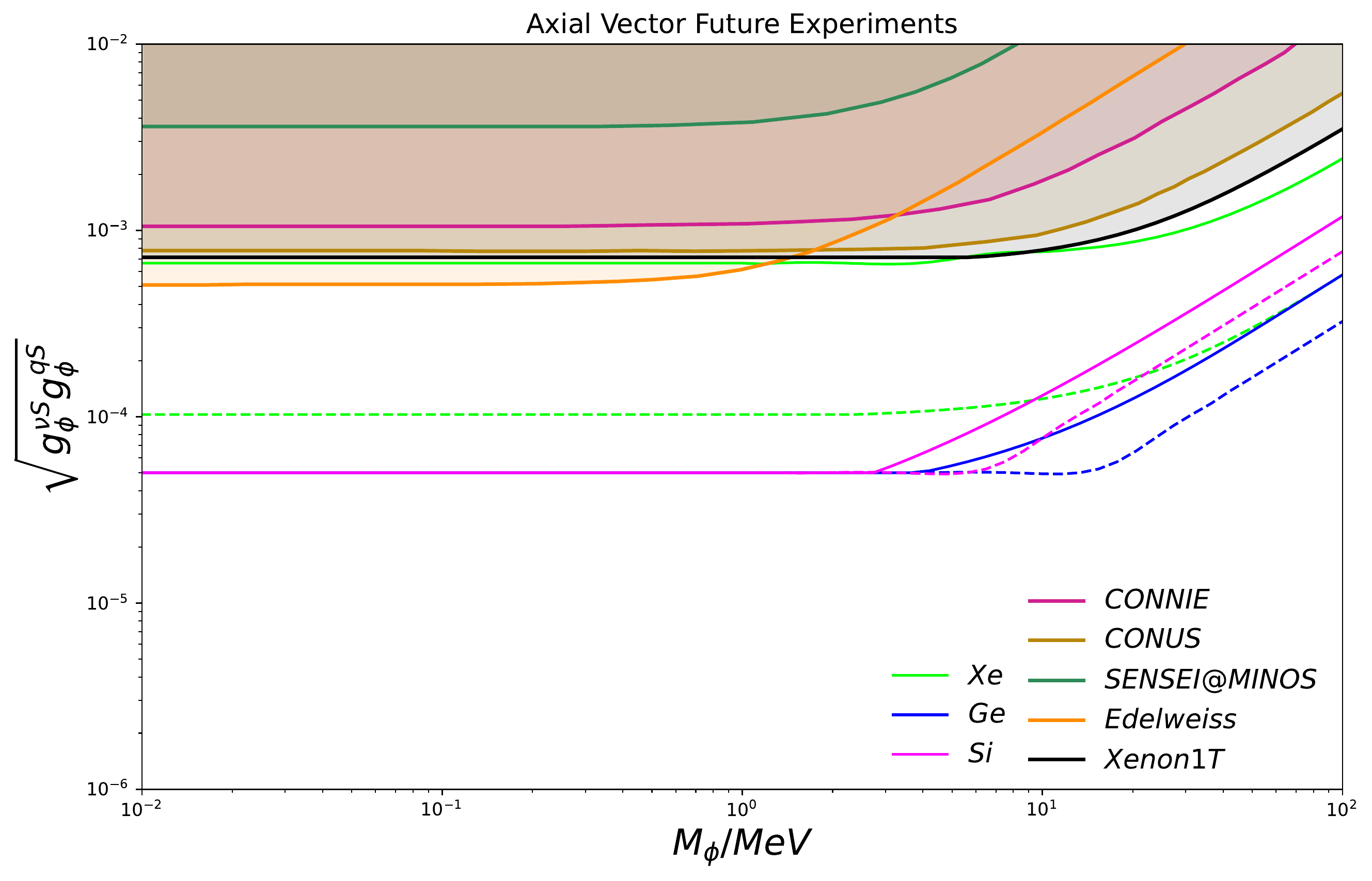}
	\caption{90$\%$ C.L. upper limits on the parameter space of the light mediators from the experimental scenarios listed in Table.~\ref{table:exp},
where the solid lines are for the nominal energy threshold and dashed lines for the optimal energy threshold. The upper, middle and lower panels are illustrated for the scalar, vector and axial vector mediators, respectively. The left and right panels are shown for the Gen-II and Future experimental scenarios, respectively. The black solid lines are the limit from the $^8$B solar neutrino results of XENON-1T. We also show constraints from the reactor and accelerator CE$\nu$NS detection, the collider searches and DM direct detection experiments.}
	\label{constraints:general}
\end{figure}
In Fig.~\ref{constraints:general} we have illustrated 90$\%$ C.L. upper limits on the parameter space of the light mediators from the experimental scenarios listed in Table.~\ref{table:exp}. The upper, middle and lower panels are illustrated for the scalar, vector and axial vector mediators, respectively.
The left and right panels are shown for the Gen-II and Future experimental scenarios, respectively.
The black solid lines are the limit from the $^8$B solar neutrino results of XENON-1T. Note that the corresponding natural abundance of isotopes has been taken into consideration.

From Fig.~\ref{constraints:general}, the constraints form solid detectors of Ge and Si are generally more stringent than liquid noble gas detectors of Xe and Ar due to the larger NP enhancement gifted by the threshold of $\mathcal{O}(10-100)$ eV. It is also obvious that improving the detector exposure suffers from severe marginal effects and increasing statistics shows very low efficiency after the constraints reach some certain levels. Though also affected by the marginal effect, better threshold, however, can generally improve the constraints with higher efficiency because all the light mediator interactions discussed in this work are suppressed by the recoil energy. In the following some key remarks are summarized for each model of the light mediators:

\begin{itemize}

\item For the scalar interaction, constraints by liquid noble gas detectors are limited beyond the level of $\sqrt{g^{\nu S}_{\phi}g^{q S}_{\phi}}>10^{-6}$. Increasing the detector exposure can hardly help because the scalar interaction contributes to little event rate beyond their thresholds. Solid detectors can give better constraints but still cannot reach $\sqrt{g^{\nu S}_{\phi}g^{q S}_{\phi}}<10^{-7}$. On the other hand, the results from XENON-1T provide a limit within the threshold band of next generation Xe detectors but can only reach with the nominal threshold and 200 t$\times$year, or with smaller exposure but better threshold. Since the scalar interaction purely enhances the event rates at a factor of $T^{-1}$, lower threshold is fundamentally important to present effective constraints on the parameter space. We also illustrate constraints on the scalar interaction from CONNIE\cite{CONNIE:2019xid}, CONUS\cite{CONUS:2021dwh} and COHERENT\cite{Corona:2022wlb}. 

\item For the vector interaction, improving the energy threshold make a difference for liquid noble gas detectors especially for the Xe detector of Gen-II.
since the vector coupling contributes to a significant cancellation at $\mathcal{O}(0.1-1)$ keV as shown in Fig.~\ref{constraints:general}, which makes the energy threshold even more important compared to the scalar and axial vector coupling.
Ar, Si and Ge detectors can constrain the vector coupling to the level of $\mathcal{O}(10^{-7}-10^{-6})$ but improving the exposure shows little effect.
While the constraints by Xe detectors are not as stringent as others due to the energy threshold, increasing statistics are more effective. 
Finally although the vector coupling mainly contributes to the event rates below 100 eV, the cancellation at $\mathcal{O}(0.1-1)$ keV is still significant to provide effective constraints with liquid noble gas detectors for an energy threshold of $\mathcal{O}(1)$ keV and enough exposure. Constraints from CONNIE\cite{CONNIE:2019xid},
COHERENT\cite{Cadeddu:2020nbr},
BaBar\cite{BaBar:2017tiz}, and NA64\cite{Banerjee:2019pds,NA64:2017vtt}
are also shown in the figure for comparison.

\item For the axial vector interaction, the enhancement of event rates is related to the abundance and nuclear spin of isotopes. In this respect, Xe detectors give better constraints compared to other nuclei since $^{131}$Xe and $^{129}$Xe, which are the isotopes of Xe with nonzero nuclear spin, have the abundance of about 21$\%$ and 26$\%$ respectively. Among the isotopes of Ge, $^{73}$Ge is the only one with a nonzero nuclear spin and the abundance is 7.7$\%$. For the similar case, $^{29}$Si has a abundance of 4.6$\%$.
As a result, increasing the exposure and threshold has little effect for improving constraints on the axial vector coupling of light mediators. Ar has no stable isotope with nonzero nuclear spin and cannot give any constraint in this case. To effectively observe the enhancement induced by the axial vector interaction, enough abundance of isotopes with nonzero nuclear spin can significantly improve the effectiveness and the enrichment of effective isotopes may compensate for the unsatisfied exposure and threshold. Constraints from these experimental scenarios also show significant advantages compared with existed constraints from CONNIE\cite{CONNIE:2019xid}, CONUS\cite{CONUS:2021dwh}, SENSEI@MINOS\cite{PhysRevLett.125.171802} and Edelweiss\cite{PhysRevLett.125.141301}.

\end{itemize}

\subsection{Constraints on the ${L_{\mu}-L_{\tau}}$ model}

The discrepancy between experimental and theoretical values of the muon anomalous magnetic dipole moment~\cite{Muong-2:2021ojo} can be explained by the the ${L_{\mu}-L_{\tau}}$ vector mediator model~\cite{Cadeddu:2021dqx,Zhou:2021vnf,Ko:2021lpx,Hapitas:2021ilr},
in which the one loop contribution to $a_{\mu}$ induced by virtual exchange of $Z^{'}$ can be written as~\cite{delta_a1,Hapitas:2021ilr}
\begin{equation}
\Delta a_{\mu}=\frac{g_{\mu\tau}^{2}}{4\pi^{2}}\int_{0}^{1} \dif z \frac{m_{\mu}^{2}z^{2}(1-z)}{m_{Z^{'}}^{2}(1-z)+m_{\mu}^{2}z^{2}}\,,
\end{equation}
where the region for $ m_{Z^{'}} $ lighter than $\sim$ 6 MeV is excluded by the big-bang nucleosynthesis (BBN) and cosmic microwave background (CMB)~\cite{constraint_cos1,constraint_cos2} and the region with $ m_{Z^{'}} $ heavier than twice the muon mass is excluded by BaBar~\cite{constraint_babar}. Thus a mass window of 5 MeV $ \lesssim $ $ m_{Z^{'}} $ $ \lesssim $ 200 MeV is still viable and can be tested using the solar neutrino CE$\nu$NS process.

\begin{figure}
	\centering
	\includegraphics[scale=0.5]{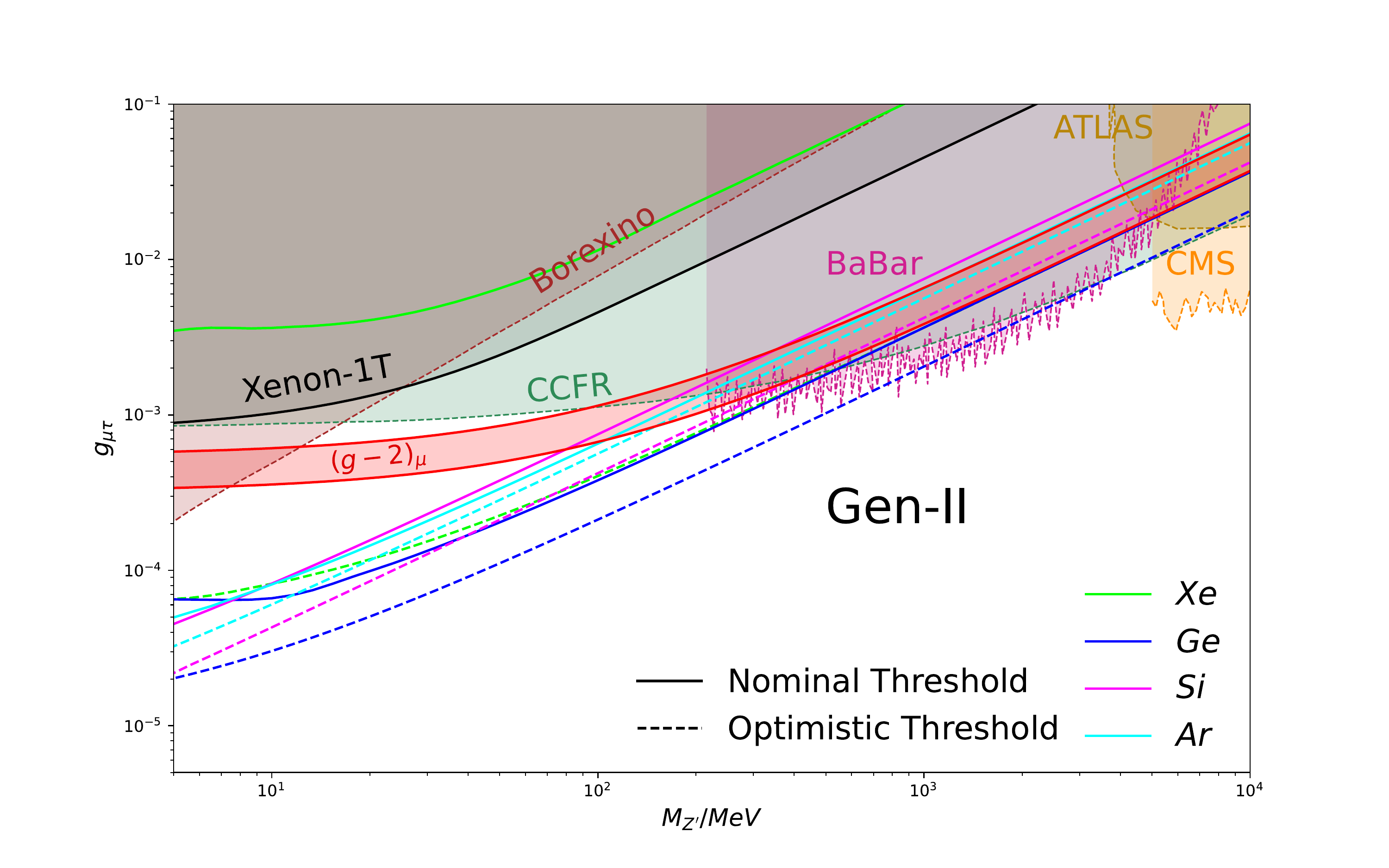}
	\vspace{-0.2cm}
	\includegraphics[scale=0.5]{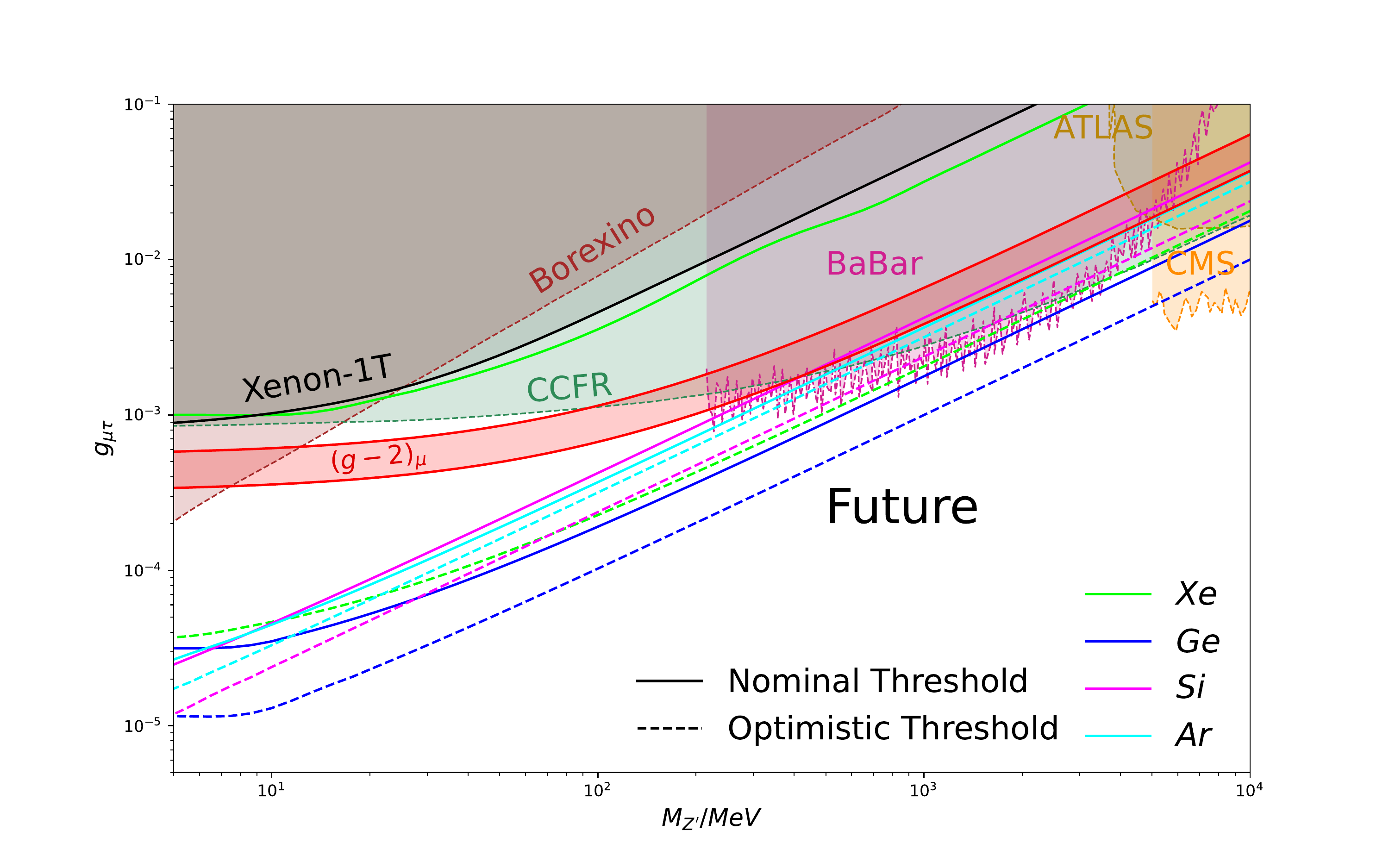}
	\vspace{-0.2cm}
	\caption{90$\%$ C.L. upper limits on the parameter space of the ${L_{\mu}-L_{\tau}}$ model from the experimental scenarios of Xe (green), Ar (blue), Si (purple) and Ge (cyan) detectors listed in Table.~\ref{table:exp}, versus the red band for the allowed range as the solution to the $ (g-2)_{\mu}$ anomaly in the ${L_{\mu}-L_{\tau}}$ model.
The solid lines are for the nominal energy threshold and dashed lines for the optimal energy threshold. The upper and lower panels are shown for the Gen-II and Future experimental scenarios, respectively. The black solid lines are the limit from the $^8$B solar neutrino results of XENON-1T. Other bounds at 95$\%$ C.L. are from BaBar~\cite{constraint_babar} (purple area), CCFR~\cite{constraint_nt} (green area), Borexino~\cite{exp_Borexino_1,exp_Borexino_2} (brown area), LHC searches in ATLAS~\cite{exp_ATLAS_1,exp_ATLAS_2} (dark yellow area), and CMS~\cite{exp_CMS} (yellow area).}
	\label{constraints:g-2}
\end{figure}

In Fig.~\ref{constraints:g-2} we have illustrated the 90$\%$ C.L. upper limits on the parameter space of the ${L_{\mu}-L_{\tau}}$ model from the experimental scenarios of Xe (green), Ar (blue), Si (purple) and Ge (cyan) detectors listed in Table.~\ref{table:exp}, versus the red band for the allowed range as the solution to the $ (g-2)_{\mu}$ anomaly in the ${L_{\mu}-L_{\tau}}$ model.
The solid lines are for the nominal energy threshold and dashed lines for the optimal energy threshold. The upper and lower panels are shown for the Gen-II and Future experimental scenarios, respectively. The black solid lines are the limit from the $^8$B solar neutrino results of XENON-1T. Other bounds at 95$\%$ C.L. are from BaBar~\cite{constraint_babar} (purple area), CCFR~\cite{constraint_nt} (green area), Borexino~\cite{exp_Borexino_1,exp_Borexino_2} (brown area), LHC searches in ATLAS~\cite{exp_ATLAS_1,exp_ATLAS_2} (dark yellow area), and CMS~\cite{exp_CMS} (yellow area).

From the figure, we can observe that most of the experimental scenarios considered in Table.~\ref{table:exp} provide excellent sensitivity on the ${L_{\mu}-L_{\tau}}$ model and can cover the allowed region of the solution to the $(g-2)_{\mu}$ anomaly with mass of $ Z^{'}$ below 100 MeV. 
Therefore, the solar neutrino CE$\nu$NS process 
is competitive for the $(g-2)_{\mu}$ exploration, and can confirm or exclude the parameter space with the mediator mass lighter than lighter than 100 MeV.
Solid detectors with Ge and Si and the Ar detectors can do even better for the mass above 100 MeV, covering almost the entire parameter space of the $(g-2)_{\mu}$ solution.
As shown in Eq.~(\ref{Q_SMmutau}), the effect induced by the $ L_{\mu}-L_{\tau}$ vector mediator is suppressed by the momentum transfer and will become significant at low recoil energy.
As a result, lower threshold detectors of Ge, Si and Ar can enhance the sensitivity to observe the deviation from the SM prediction and make the constraints reach a better level. 
The Xe detectors cannot give competitive constraints with the nominal threshold, but improving the threshold can significantly reach the parameter space of the $(g-2)_{\mu}$ solution. In contrast, a improvement of the detector exposure cannot largely increase the sensitivity. Finally we would like to remark that although the current limit from the $^8$B solar neutrino results of XENON-1T cannot rule out the $ L_{\mu}-L_{\tau}$ solution to the $ (g-2)_{\mu} $ anomaly, it proved the feasibility of the same method in future DM direct detection experiments.

\section{Conclusion}

Dark matter (DM) direct detection experiments are entering the multiple-ton era and will be sensitive to the CE$\nu$NS process of solar neutrinos, enabling the possibility to explore contributions from new physics with light mediators at the low energy range. In this work we have explored three flavor-universal light mediator models (scalar, vector and axial vector) and the corresponding contributions to the solar neutrino CE$\nu$NS process.
Motivated by the current status and future plan of the DM direct detection experiments, we have presented the sensitivity of light mediators from different nuclear targets and detector techniques. We have shown that the detector energy threshold and exposure are crucial parameters that significant affect the levels of the sensitivity.
The constraints from the $^8$B solar neutrino CE$\nu$NS measurements of XENON-1T are also derived, which have proved the feasibility and power of the future measurement with the same method.
Finally, We have illustrated that the solar neutrino CE$\nu$NS process can provide stringent limitation on the $ L_{\mu}-L_{\tau} $ model with the vector mediator mass below 100 MeV, which covers the viable parameter space of the solution to the $ (g-2)_{\mu}$ anomaly, but is completely not accessible by other probes. We encourage the current and future DM direct detection experiments pursue this important goal with dedicated efforts on the critical energy threshold and exposure.

\section*{Acknowledgements}
The authors are grateful to Dr.~Yiyu Zhang for helpful discussions. The work of YFL and SYX was supported by National Natural Science Foundation of China under Grant Nos.~12075255, 12075254, 11775231 and 11835013, by Beijing Natural Science Foundation under Grant No.~1192019, by the Key Research Program of the Chinese Academy of Sciences under Grant No.~XDPB15.
YFL is also grateful for the support by the CAS Center for Excellence in Particle Physics.


\bibliographystyle{h-physrev5}
\bibliography{BIB}

\begin{thebibliography}{100}

\bibitem{Freedman:1973yd}
D.~Z. Freedman, {Coherent Neutrino Nucleus Scattering as a Probe of the Weak
  Neutral Current},
\newblock Phys. Rev. D {\bf 9}, 1389 (1974).

\bibitem{Freedman:1977xn}
D.~Z. Freedman, D.~N. Schramm, and D.~L. Tubbs, {The Weak Neutral Current and
  Its Effects in Stellar Collapse},
\newblock Ann. Rev. Nucl. Part. Sci. {\bf 27}, 167 (1977).

\bibitem{COHERENT:2017ipa}
COHERENT, D.~Akimov {\em et~al.}, {Observation of Coherent Elastic
  Neutrino-Nucleus Scattering},
\newblock Science {\bf 357}, 1123 (2017), arXiv:1708.01294.

\bibitem{COHERENT:2020iec}
COHERENT, D.~Akimov {\em et~al.}, {First Measurement of Coherent Elastic
  Neutrino-Nucleus Scattering on Argon},
\newblock Phys. Rev. Lett. {\bf 126}, 012002 (2021), arXiv:2003.10630.

\bibitem{Akimov:2021dab}
D.~Akimov {\em et~al.}, {Measurement of the Coherent Elastic Neutrino-Nucleus
  Scattering Cross Section on CsI by COHERENT},
\newblock (2021), arXiv:2110.07730.

\bibitem{Cadeddu:2017etk}
M.~Cadeddu, C.~Giunti, Y.~F. Li, and Y.~Y. Zhang, {Average CsI neutron density
  distribution from COHERENT data},
\newblock Phys. Rev. Lett. {\bf 120}, 072501 (2018), arXiv:1710.02730.

\bibitem{Cadeddu:2020lky}
M.~Cadeddu {\em et~al.}, {Physics results from the first COHERENT observation
  of coherent elastic neutrino-nucleus scattering in argon and their
  combination with cesium-iodide data},
\newblock Phys. Rev. D {\bf 102}, 015030 (2020), arXiv:2005.01645.

\bibitem{Cadeddu:2021ijh}
M.~Cadeddu {\em et~al.}, {New insights into nuclear physics and weak mixing
  angle using electroweak probes},
\newblock (2021), arXiv:2102.06153.

\bibitem{Ciuffoli:2018qem}
E.~Ciuffoli, J.~Evslin, Q.~Fu, and J.~Tang, {Extracting nuclear form factors
  with coherent neutrino scattering},
\newblock Phys. Rev. D {\bf 97}, 113003 (2018), arXiv:1801.02166.

\bibitem{Papoulias:2019lfi}
D.~K. Papoulias, T.~S. Kosmas, R.~Sahu, V.~K.~B. Kota, and M.~Hota,
  {Constraining nuclear physics parameters with current and future COHERENT
  data},
\newblock Phys. Lett. B {\bf 800}, 135133 (2020), arXiv:1903.03722.

\bibitem{Coloma:2020nhf}
P.~Coloma, I.~Esteban, M.~C. Gonzalez-Garcia, and J.~Menendez, {Determining the
  nuclear neutron distribution from Coherent Elastic neutrino-Nucleus
  Scattering: current results and future prospects},
\newblock JHEP {\bf 08}, 030 (2020), arXiv:2006.08624.

\bibitem{Papoulias:2017qdn}
D.~K. Papoulias and T.~S. Kosmas, {COHERENT constraints to conventional and
  exotic neutrino physics},
\newblock Phys. Rev. D {\bf 97}, 033003 (2018), arXiv:1711.09773.

\bibitem{Cadeddu:2018izq}
M.~Cadeddu and F.~Dordei, {Reinterpreting the weak mixing angle from atomic
  parity violation in view of the Cs neutron rms radius measurement from
  COHERENT},
\newblock Phys. Rev. D {\bf 99}, 033010 (2019), arXiv:1808.10202.

\bibitem{Cadeddu:2018dux}
M.~Cadeddu {\em et~al.}, {Neutrino Charge Radii from COHERENT Elastic
  Neutrino-Nucleus Scattering},
\newblock Phys. Rev. D {\bf 98}, 113010 (2018), arXiv:1810.05606,
\newblock [Erratum: Phys.Rev.D 101, 059902 (2020)].

\bibitem{Cadeddu:2019eta}
M.~Cadeddu, F.~Dordei, C.~Giunti, Y.~F. Li, and Y.~Y. Zhang, {Neutrino,
  electroweak, and nuclear physics from COHERENT elastic neutrino-nucleus
  scattering with refined quenching factor},
\newblock Phys. Rev. D {\bf 101}, 033004 (2020), arXiv:1908.06045.

\bibitem{Kim:2021lun}
J.~E. Kim, A.~Dasgupta, and S.~K. Kang, {Probing Neutrino Dipole Portal at
  COHERENT Experiment},
\newblock (2021), arXiv:2108.12998.

\bibitem{Coloma:2017ncl}
P.~Coloma, M.~C. Gonzalez-Garcia, M.~Maltoni, and T.~Schwetz, {COHERENT
  Enlightenment of the Neutrino Dark Side},
\newblock Phys. Rev. D {\bf 96}, 115007 (2017), arXiv:1708.02899.

\bibitem{Liao:2017uzy}
J.~Liao and D.~Marfatia, {COHERENT constraints on nonstandard neutrino
  interactions},
\newblock Phys. Lett. B {\bf 775}, 54 (2017), arXiv:1708.04255.

\bibitem{Denton:2018xmq}
P.~B. Denton, Y.~Farzan, and I.~M. Shoemaker, {Testing large non-standard
  neutrino interactions with arbitrary mediator mass after COHERENT data},
\newblock JHEP {\bf 07}, 037 (2018), arXiv:1804.03660.

\bibitem{AristizabalSierra:2018eqm}
D.~Aristizabal~Sierra, V.~De~Romeri, and N.~Rojas, {COHERENT analysis of
  neutrino generalized interactions},
\newblock Phys. Rev. D {\bf 98}, 075018 (2018), arXiv:1806.07424.

\bibitem{Giunti:2019xpr}
C.~Giunti, {General COHERENT constraints on neutrino nonstandard interactions},
\newblock Phys. Rev. D {\bf 101}, 035039 (2020), arXiv:1909.00466.

\bibitem{Farzan:2018gtr}
Y.~Farzan, M.~Lindner, W.~Rodejohann, and X.-J. Xu, {Probing neutrino coupling
  to a light scalar with coherent neutrino scattering},
\newblock JHEP {\bf 05}, 066 (2018), arXiv:1802.05171.

\bibitem{Abdullah:2018ykz}
M.~Abdullah {\em et~al.}, {Coherent elastic neutrino nucleus scattering as a
  probe of a Z' through kinetic and mass mixing effects},
\newblock Phys. Rev. D {\bf 98}, 015005 (2018), arXiv:1803.01224.

\bibitem{Ge:2017mcq}
S.-F. Ge and I.~M. Shoemaker, {Constraining Photon Portal Dark Matter with
  Texono and Coherent Data},
\newblock JHEP {\bf 11}, 066 (2018), arXiv:1710.10889.

\bibitem{Cadeddu:2020nbr}
M.~Cadeddu {\em et~al.}, {Constraints on light vector mediators through
  coherent elastic neutrino nucleus scattering data from COHERENT},
\newblock JHEP {\bf 01}, 116 (2021), arXiv:2008.05022.

\bibitem{Boehm:2018sux}
C.~B\oe{}hm {\em et~al.}, {How high is the neutrino floor?},
\newblock JCAP {\bf 01}, 043 (2019), arXiv:1809.06385.

\bibitem{Gonzalez-Garcia:2018dep}
M.~C. Gonzalez-Garcia, M.~Maltoni, Y.~F. Perez-Gonzalez, and
  R.~Zukanovich~Funchal, {Neutrino Discovery Limit of Dark Matter Direct
  Detection Experiments in the Presence of Non-Standard Interactions},
\newblock JHEP {\bf 07}, 019 (2018), arXiv:1803.03650.

\bibitem{Papoulias:2018uzy}
D.~K. Papoulias, R.~Sahu, T.~S. Kosmas, V.~K.~B. Kota, and B.~Nayak, {Novel
  neutrino-floor and dark matter searches with deformed shell model
  calculations},
\newblock Adv. High Energy Phys. {\bf 2018}, 6031362 (2018), arXiv:1804.11319.

\bibitem{Chao:2019pyh}
W.~Chao, J.-G. Jiang, X.~Wang, and X.-Y. Zhang, {Direct Detections of Dark
  Matter in the Presence of Non-standard Neutrino Interactions},
\newblock JCAP {\bf 08}, 010 (2019), arXiv:1904.11214.

\bibitem{AristizabalSierra:2021kht}
D.~Aristizabal~Sierra, V.~De~Romeri, L.~J. Flores, and D.~K. Papoulias, {Impact
  of COHERENT measurements, cross section uncertainties and new interactions on
  the neutrino floor},
\newblock (2021), arXiv:2109.03247.

\bibitem{OHare:2020lva}
C.~A.~J. O'Hare, {Can we overcome the neutrino floor at high masses?},
\newblock Phys. Rev. D {\bf 102}, 063024 (2020), arXiv:2002.07499.

\bibitem{Lang:2016zhv}
R.~F. Lang, C.~McCabe, S.~Reichard, M.~Selvi, and I.~Tamborra, {Supernova
  neutrino physics with xenon dark matter detectors: A timely perspective},
\newblock Phys. Rev. D {\bf 94}, 103009 (2016), arXiv:1606.09243.

\bibitem{Pattavina:2020cqc}
L.~Pattavina, N.~Ferreiro~Iachellini, and I.~Tamborra, {Neutrino observatory
  based on archaeological lead},
\newblock Phys. Rev. D {\bf 102}, 063001 (2020), arXiv:2004.06936.

\bibitem{RES-NOVA:2021gqp}
RES-NOVA, L.~Pattavina {\em et~al.}, {RES-NOVA sensitivity to core-collapse and
  failed core-collapse supernova neutrinos},
\newblock JCAP {\bf 10}, 064 (2021), arXiv:2103.08672.

\bibitem{Raj:2019sci}
N.~Raj, {Neutrinos from Type Ia and failed core-collapse supernovae at dark
  matter detectors},
\newblock Phys. Rev. Lett. {\bf 124}, 141802 (2020), arXiv:1907.05533.

\bibitem{Raj:2019wpy}
N.~Raj, V.~Takhistov, and S.~J. Witte, {Presupernova neutrinos in large dark
  matter direct detection experiments},
\newblock Phys. Rev. D {\bf 101}, 043008 (2020), arXiv:1905.09283.

\bibitem{Huang:2021enl}
X.-R. Huang, S.~Zha, and L.-W. Chen, {Supernova Preshock Neutronization Burst
  as a Probe of Non-Standard Neutrino Interactions},
\newblock (2021), arXiv:2110.07249.

\bibitem{Munoz:2021sad}
V.~Munoz, V.~Takhistov, S.~J. Witte, and G.~M. Fuller, {Exploring the origin of
  supermassive black holes with coherent neutrino scattering},
\newblock JCAP {\bf 11}, 020 (2021), arXiv:2102.00885.

\bibitem{Calabrese:2021zfq}
R.~Calabrese, D.~F.~G. Fiorillo, G.~Miele, S.~Morisi, and A.~Palazzo,
  {Primordial Black Hole Dark Matter evaporating on the Neutrino Floor},
\newblock (2021), arXiv:2106.02492.

\bibitem{Barbeau:2021exu}
P.~S. Barbeau, Y.~Efremenko, and K.~Scholberg, {COHERENT at the Spallation
  Neutron Source},
\newblock (2021), arXiv:2111.07033.

\bibitem{Baxter:2019mcx}
D.~Baxter {\em et~al.}, {Coherent Elastic Neutrino-Nucleus Scattering at the
  European Spallation Source},
\newblock JHEP {\bf 02}, 123 (2020), arXiv:1911.00762.

\bibitem{CONUS:2020skt}
CONUS, H.~Bonet {\em et~al.}, {Constraints on Elastic Neutrino Nucleus
  Scattering in the Fully Coherent Regime from the CONUS Experiment},
\newblock Phys. Rev. Lett. {\bf 126}, 041804 (2021), arXiv:2011.00210.

\bibitem{CONUS:2021dwh}
CONUS, H.~Bonet {\em et~al.}, {Novel constraints on neutrino physics beyond the
  standard model from the CONUS experiment},
\newblock (2021), arXiv:2110.02174.

\bibitem{CONNIE:2019xid}
CONNIE, A.~Aguilar-Arevalo {\em et~al.}, {Search for light mediators in the
  low-energy data of the CONNIE reactor neutrino experiment},
\newblock JHEP {\bf 04}, 054 (2020), arXiv:1910.04951.

\bibitem{CONNIE:2021ngo}
CONNIE, A.~Aguilar-Arevalo {\em et~al.}, {Search for coherent elastic
  neutrino-nucleus scattering at a nuclear reactor with CONNIE 2019 data},
\newblock (2021), arXiv:2110.13033.

\bibitem{Davis:1968cp}
R.~Davis, Jr., D.~S. Harmer, and K.~C. Hoffman, {Search for neutrinos from the
  sun},
\newblock Phys. Rev. Lett. {\bf 20}, 1205 (1968).

\bibitem{GALLEX:1992gcp}
GALLEX, P.~Anselmann {\em et~al.}, {Solar neutrinos observed by GALLEX at Gran
  Sasso.},
\newblock Phys. Lett. B {\bf 285}, 376 (1992).

\bibitem{GNO:2000avz}
GNO, M.~Altmann {\em et~al.}, {GNO solar neutrino observations: Results for GNO
  I},
\newblock Phys. Lett. B {\bf 490}, 16 (2000), arXiv:hep-ex/0006034.

\bibitem{Abazov:1991rx}
A.~I. Abazov {\em et~al.}, {Search for neutrinos from sun using the reaction
  Ga-71 (electron-neutrino e-) Ge-71},
\newblock Phys. Rev. Lett. {\bf 67}, 3332 (1991).

\bibitem{SAGE:1999uje}
SAGE, J.~N. Abdurashitov {\em et~al.}, Measurement of the solar neutrino
  capture rate by sage and implications for neutrino oscillations in vacuum,
\newblock Phys. Rev. Lett. {\bf 83}, 4686 (1999), arXiv:astro-ph/9907131.

\bibitem{Cleveland:1998nv}
Homestake, B.~T. Cleveland {\em et~al.}, Measurement of the solar electron
  neutrino flux with the homestake chlorine detector,
\newblock Astrophys. J. {\bf 496}, 505 (1998).

\bibitem{SNO:2001kpb}
SNO, Q.~R. Ahmad {\em et~al.}, Measurement of the rate of $\nu_e + d \to p + p
  + e^-$ interactions produced by $^8$b solar neutrinos at the sudbury neutrino
  observatory,
\newblock Phys. Rev. Lett. {\bf 87}, 071301 (2001), arXiv:nucl-ex/0106015.

\bibitem{Kamiokande-II:1989hkh}
Kamiokande, K.~S. Hirata {\em et~al.}, Observation of b-8 solar neutrinos in
  the kamiokande-ii detector,
\newblock Phys. Rev. Lett. {\bf 63}, 16 (1989).

\bibitem{Super-Kamiokande:2001ljr}
Super-Kamiokande, S.~Fukuda {\em et~al.}, Solar $^8\mathrm{B}$ and $hep$
  neutrino measurements from 1258 days of super-kamiokande data,
\newblock Phys. Rev. Lett. {\bf 86}, 5651 (2001), arXiv:hep-ex/0103032.

\bibitem{Borexino:2007kvk}
Borexino, C.~Arpesella {\em et~al.}, First real time detection of
  ${}^{7}\text{B}$ solar neutrinos by borexino,
\newblock Phys. Lett. {\bf B658}, 101 (2008), arXiv:0708.2251.

\bibitem{Borexino:2008fkj}
Borexino, G.~Bellini {\em et~al.}, Measurement of the solar $^{8}\text{B}$
  neutrino flux with 246 live days of borexino and observation of the msw
  vacuum-matter transition,
\newblock Phys. Rev. {\bf D82}, 033006 (2010), arXiv:0808.2868.

\bibitem{Collaboration:2011nga}
Borexino, G.~Bellini {\em et~al.}, First evidence of pep solar neutrinos by
  direct detection in borexino,
\newblock Phys. Rev. Lett. {\bf 108}, 051302 (2012), arXiv:1110.3230.

\bibitem{BOREXINO:2014pcl}
Borexino, G.~Bellini {\em et~al.}, Neutrinos from the primary proton-proton
  fusion process in the sun,
\newblock Nature {\bf 512}, 383 (2014).

\bibitem{BOREXINO:2020aww}
Borexino, M.~Agostini {\em et~al.}, First direct experimental evidence of cno
  neutrinos,
\newblock Nature {\bf 587}, 577 (2020), arXiv:2006.15115.

\bibitem{SNO:2002tuh}
SNO, Q.~R. Ahmad {\em et~al.}, Direct evidence for neutrino flavor
  transformation from neutral-current interactions in the sudbury neutrino
  observatory,
\newblock Phys. Rev. Lett. {\bf 89}, 011301 (2002), arXiv:nucl-ex/0204008.

\bibitem{SNO:2003bmh}
SNO, S.~N. Ahmed {\em et~al.}, Measurement of the total active 8b solar
  neutrino flux at the sudbury neutrino observatory with enhanced neutral
  current sensitivity,
\newblock Phys. Rev. Lett. {\bf 92}, 181301 (2004), arXiv:nucl-ex/0309004.

\bibitem{SNO:2008gqy}
SNO, B.~Aharmim {\em et~al.}, An independent measurement of the total active 8b
  solar neutrino flux using an array of 3he proportional counters at the
  sudbury neutrino observatory,
\newblock Phys. Rev. Lett. {\bf 101}, 111301 (2008), arXiv:0806.0989.

\bibitem{Bahcall:1998wm}
J.~N. Bahcall, S.~Basu, and M.~H. Pinsonneault, How uncertain are solar
  neutrino predictions?,
\newblock Phys. Lett. {\bf B433}, 1 (1998), arXiv:astro-ph/9805135.

\bibitem{Bahcall:2000nu}
J.~N. Bahcall, M.~H. Pinsonneault, and S.~Basu, Solar models: Current epoch and
  time dependences, neutrinos, and helioseismological properties,
\newblock Astrophys. J. {\bf 555}, 990 (2001), arXiv:astro-ph/0010346.

\bibitem{Bahcall:2004fg}
J.~N. Bahcall and M.~H. Pinsonneault, What do we (not) know theoretically about
  solar neutrino fluxes?,
\newblock Phys. Rev. Lett. {\bf 92}, 121301 (2004), arXiv:astro-ph/0402114.

\bibitem{Vinyoles:2016djt}
N.~Vinyoles {\em et~al.}, {A new Generation of Standard Solar Models},
\newblock Astrophys. J. {\bf 835}, 202 (2017), arXiv:1611.09867.

\bibitem{Dutta:2019oaj}
B.~Dutta and L.~E. Strigari, {Neutrino physics with dark matter detectors},
\newblock Ann. Rev. Nucl. Part. Sci. {\bf 69}, 137 (2019), arXiv:1901.08876.

\bibitem{Billard:2014yka}
J.~Billard, L.~E. Strigari, and E.~Figueroa-Feliciano, {Solar neutrino physics
  with low-threshold dark matter detectors},
\newblock Phys. Rev. D {\bf 91}, 095023 (2015), arXiv:1409.0050.

\bibitem{Cerdeno:2016sfi}
D.~G. Cerde\~no {\em et~al.}, {Physics from solar neutrinos in dark matter
  direct detection experiments},
\newblock JHEP {\bf 05}, 118 (2016), arXiv:1604.01025,
\newblock [Erratum: JHEP 09, 048 (2016)].

\bibitem{Harnik:2012ni}
R.~Harnik, J.~Kopp, and P.~A.~N. Machado, {Exploring nu Signals in Dark Matter
  Detectors},
\newblock JCAP {\bf 07}, 026 (2012), arXiv:1202.6073.

\bibitem{Billard:2021uyg}
J.~Billard {\em et~al.}, {Direct Detection of Dark Matter -- APPEC Committee
  Report},
\newblock (2021), arXiv:2104.07634.

\bibitem{Liu:2017drf}
J.~Liu, X.~Chen, and X.~Ji, {Current status of direct dark matter detection
  experiments},
\newblock Nature Phys. {\bf 13}, 212 (2017), arXiv:1709.00688.

\bibitem{Schumann:2019eaa}
M.~Schumann, {Direct Detection of WIMP Dark Matter: Concepts and Status},
\newblock J. Phys. G {\bf 46}, 103003 (2019), arXiv:1903.03026.

\bibitem{PandaX:2018wtu}
PandaX, H.~Zhang {\em et~al.}, {Dark matter direct search sensitivity of the
  PandaX-4T experiment},
\newblock Sci. China Phys. Mech. Astron. {\bf 62}, 31011 (2019),
  arXiv:1806.02229.

\bibitem{XENON:2020kmp}
XENON, E.~Aprile {\em et~al.}, {Projected WIMP sensitivity of the XENONnT dark
  matter experiment},
\newblock JCAP {\bf 11}, 031 (2020), arXiv:2007.08796.

\bibitem{LZ:2019sgr}
LZ, D.~S. Akerib {\em et~al.}, {The LUX-ZEPLIN (LZ) Experiment},
\newblock Nucl. Instrum. Meth. A {\bf 953}, 163047 (2020), arXiv:1910.09124.

\bibitem{DARWIN:2016hyl}
DARWIN, J.~Aalbers {\em et~al.}, {DARWIN: towards the ultimate dark matter
  detector},
\newblock JCAP {\bf 11}, 017 (2016), arXiv:1606.07001.

\bibitem{DarkSide-20k:2017zyg}
DarkSide-20k, C.~E. Aalseth {\em et~al.}, {DarkSide-20k: A 20 tonne two-phase
  LAr TPC for direct dark matter detection at LNGS},
\newblock Eur. Phys. J. Plus {\bf 133}, 131 (2018), arXiv:1707.08145.

\bibitem{PandaX-4T:2021bab}
PandaX-4T, Y.~Meng {\em et~al.}, {Dark Matter Search Results from the PandaX-4T
  Commissioning Run},
\newblock (2021), arXiv:2107.13438.

\bibitem{SuperCDMS:2016wui}
SuperCDMS, R.~Agnese {\em et~al.}, {Projected Sensitivity of the SuperCDMS
  SNOLAB experiment},
\newblock Phys. Rev. D {\bf 95}, 082002 (2017), arXiv:1610.00006.

\bibitem{EDELWEISS:2017uga}
EDELWEISS, Q.~Arnaud {\em et~al.}, {Optimizing EDELWEISS detectors for low-mass
  WIMP searches},
\newblock Phys. Rev. D {\bf 97}, 022003 (2018), arXiv:1707.04308.

\bibitem{XENON:2020gfr}
XENON, E.~Aprile {\em et~al.}, {Search for Coherent Elastic Scattering of Solar
  $^8$B Neutrinos in the XENON1T Dark Matter Experiment},
\newblock Phys. Rev. Lett. {\bf 126}, 091301 (2021), arXiv:2012.02846.

\bibitem{Muong-2:2021ojo}
Muon g-2, B.~Abi {\em et~al.}, {Measurement of the Positive Muon Anomalous
  Magnetic Moment to 0.46 ppm},
\newblock Phys. Rev. Lett. {\bf 126}, 141801 (2021), arXiv:2104.03281.

\bibitem{Muong-2:2006rrc}
Muon g-2, G.~W. Bennett {\em et~al.}, {Final Report of the Muon E821 Anomalous
  Magnetic Moment Measurement at BNL},
\newblock Phys. Rev. D {\bf 73}, 072003 (2006), arXiv:hep-ex/0602035.

\bibitem{Aoyama:2020ynm}
T.~Aoyama {\em et~al.}, {The anomalous magnetic moment of the muon in the
  Standard Model},
\newblock Phys. Rept. {\bf 887}, 1 (2020), arXiv:2006.04822.

\bibitem{Athron:2021iuf}
P.~Athron {\em et~al.}, {New physics explanations of $a_\mu$ in light of the
  FNAL muon $g-2$ measurement},
\newblock JHEP {\bf 09}, 080 (2021), arXiv:2104.03691.

\bibitem{Lindner:2016bgg}
M.~Lindner, M.~Platscher, and F.~S. Queiroz, {A Call for New Physics : The Muon
  Anomalous Magnetic Moment and Lepton Flavor Violation},
\newblock Phys. Rept. {\bf 731}, 1 (2018), arXiv:1610.06587.

\bibitem{Cadeddu:2021dqx}
M.~Cadeddu, N.~Cargioli, F.~Dordei, C.~Giunti, and E.~Picciau, {Muon and
  electron g-2 and proton and cesium weak charges implications on dark Zd
  models},
\newblock Phys. Rev. D {\bf 104}, 011701 (2021), arXiv:2104.03280.

\bibitem{Zhou:2021vnf}
S.~Zhou, {Neutrino Masses, Leptonic Flavor Mixing and Muon $(g-2)$ in the
  Seesaw Model with the $U(1)^{}_{L^{}_\mu-L^{}_\tau}$ Gauge Symmetry},
\newblock (2021), arXiv:2104.06858.

\bibitem{Ko:2021lpx}
P.~Ko, T.~Nomura, and H.~Okada, {Muon $g-2$, $B\to K^{(*)}\mu^+ \mu^-$
  anomalies, and leptophilic dark matter in $U(1)_{\mu-\tau}$ gauge symmetry},
\newblock (2021), arXiv:2110.10513.

\bibitem{Hapitas:2021ilr}
T.~Hapitas, D.~Tuckler, and Y.~Zhang, {General Kinetic Mixing in Gauged
  $U(1)_{L_\mu-L_\tau}$ Model for Muon $g-2$ and Dark Matter},
\newblock (2021), arXiv:2108.12440.

\bibitem{Drukier:1984vhf}
A.~Drukier and L.~Stodolsky, {Principles and Applications of a Neutral Current
  Detector for Neutrino Physics and Astronomy},
\newblock Phys. Rev. D {\bf 30}, 2295 (1984).

\bibitem{Barranco:2005yy}
J.~Barranco, O.~G. Miranda, and T.~I. Rashba, {Probing new physics with
  coherent neutrino scattering off nuclei},
\newblock JHEP {\bf 12}, 021 (2005), arXiv:hep-ph/0508299.

\bibitem{Patton:2012jr}
K.~Patton, J.~Engel, G.~C. McLaughlin, and N.~Schunck, {Neutrino-nucleus
  coherent scattering as a probe of neutron density distributions},
\newblock Phys. Rev. C {\bf 86}, 024612 (2012), arXiv:1207.0693.

\bibitem{PhysRevLett.125.141301}
EDELWEISS Collaboration, Q.~Arnaud {\em et~al.}, First germanium-based
  constraints on sub-mev dark matter with the edelweiss experiment,
\newblock Phys. Rev. Lett. {\bf 125}, 141301 (2020).

\bibitem{Helm:1956zz}
R.~H. Helm, {Inelastic and Elastic Scattering of 187-Mev Electrons from
  Selected Even-Even Nuclei},
\newblock Phys. Rev. {\bf 104}, 1466 (1956).

\bibitem{Cirelli:2013ufw}
M.~Cirelli, E.~Del~Nobile, and P.~Panci, {Tools for model-independent bounds in
  direct dark matter searches},
\newblock JCAP {\bf 10}, 019 (2013), arXiv:1307.5955.

\bibitem{AristizabalSierra:2019ykk}
D.~Aristizabal~Sierra, B.~Dutta, S.~Liao, and L.~E. Strigari, {Coherent elastic
  neutrino-nucleus scattering in multi-ton scale dark matter experiments:
  Classification of vector and scalar interactions new physics signals},
\newblock JHEP {\bf 12}, 124 (2019), arXiv:1910.12437.

\bibitem{Ellis:2018dmb}
J.~Ellis, N.~Nagata, and K.~A. Olive, {Uncertainties in WIMP Dark Matter
  Scattering Revisited},
\newblock Eur. Phys. J. C {\bf 78}, 569 (2018), arXiv:1805.09795.

\bibitem{Hoferichter:2015dsa}
M.~Hoferichter, J.~Ruiz~de Elvira, B.~Kubis, and U.-G. Mei\ss{}ner,
  {High-Precision Determination of the Pion-Nucleon \ensuremath{\sigma} Term
  from Roy-Steiner Equations},
\newblock Phys. Rev. Lett. {\bf 115}, 092301 (2015), arXiv:1506.04142.

\bibitem{Bertuzzo:2017tuf}
E.~Bertuzzo, F.~F. Deppisch, S.~Kulkarni, Y.~F. Perez~Gonzalez, and
  R.~Zukanovich~Funchal, {Dark Matter and Exotic Neutrino Interactions in
  Direct Detection Searches},
\newblock JHEP {\bf 04}, 073 (2017), arXiv:1701.07443.

\bibitem{XENON100:2010cgk}
XENON100, E.~Aprile {\em et~al.}, First dark matter results from the xenon100
  experiment,
\newblock Phys. Rev. Lett. {\bf 105}, 131302 (2010), arXiv:1005.0380.

\bibitem{LUX:2013afz}
LUX, D.~Akerib {\em et~al.}, First results from the lux dark matter experiment
  at the sanford underground research facility,
\newblock Phys. Rev. Lett. {\bf 112}, 091303 (2014), arXiv:1310.8214.

\bibitem{PandaX-II:2016vec}
PandaX-II, A.~Tan {\em et~al.}, {Dark Matter Results from First 98.7 Days of
  Data from the PandaX-II Experiment},
\newblock Phys. Rev. Lett. {\bf 117}, 121303 (2016), arXiv:1607.07400.

\bibitem{BKar}
DarkSide, P.~Agnes {\em et~al.}, {DarkSide-50 532-day Dark Matter Search with
  Low-Radioactivity Argon},
\newblock Phys. Rev. D {\bf 98}, 102006 (2018), arXiv:1802.07198.

\bibitem{DEAP3600}
DEAP Collaboration, {DEAP Collaboration}, Search for dark matter with a 231-day
  exposure of liquid argon using deap-3600 at snolab,
\newblock Phys. Rev. D {\bf 100}, 022004 (2019).

\bibitem{CDMS:2013juh}
CDMS, R.~Agnese {\em et~al.}, Dark matter search results using the silicon
  detectors of cdms ii,
\newblock Phys. Rev. Lett. {\bf 111}, 251301 (2013), arXiv:1304.4279.

\bibitem{SuperCDMS:2014cds}
SuperCDMS, R.~Agnese {\em et~al.}, Search for low-mass wimps with supercdms,
\newblock Phys. Rev. Lett. {\bf 112}, 241302 (2014), arXiv:1402.7137.

\bibitem{Aramaki:2016spe}
SuperCDMS, T.~Aramaki, {Recent results from the second CDMSlite run and
  overview of the SuperCDMS SNOLAB project},
\newblock PoS {\bf DSU2015}, 030 (2016).

\bibitem{CDEX:2018lau}
H.~Jiang {\em et~al.}, Limits on light wimps from the first 102.8 kg-days data
  of the cdex-10 experiment,
\newblock Phys.Rev.Lett. {\bf 120}, 241301 (2018), arXiv:1802.09016.

\bibitem{DAMIC:2021esz}
DAMIC, M.~Traina {\em et~al.}, {Results on low-mass weakly interacting massive
  particles from a 11 kg d target exposure of DAMIC at SNOLAB},
\newblock PoS {\bf ICRC2021}, 539 (2021), arXiv:2108.05983.

\bibitem{SENSEI:2020dpa}
SENSEI, L.~Barak {\em et~al.}, {SENSEI: Direct-Detection Results on sub-GeV
  Dark Matter from a New Skipper-CCD},
\newblock Phys. Rev. Lett. {\bf 125}, 171802 (2020), arXiv:2004.11378.

\bibitem{Tiffenberg:2017aac}
SENSEI, J.~Tiffenberg {\em et~al.}, {Single-electron and single-photon
  sensitivity with a silicon Skipper CCD},
\newblock Phys. Rev. Lett. {\bf 119}, 131802 (2017), arXiv:1706.00028.

\bibitem{BKxe}
DARWIN, C.~Macolino, {DARWIN: direct dark matter search with the ultimate
  detector},
\newblock J. Phys. Conf. Ser. {\bf 1468}, 012068 (2020).

\bibitem{Xenon}
XENON Collaboration, {XENON Collaboration}, Search for coherent elastic
  scattering of solar $^{8}\mathrm{B}$ neutrinos in the xenon1t dark matter
  experiment,
\newblock Phys. Rev. Lett. {\bf 126}, 091301 (2021).

\bibitem{Corona:2022wlb}
M.~A. Corona {\em et~al.}, {Probing light mediators and $(g-2)_{\mu}$ through
  detection of coherent elastic neutrino nucleus scattering at COHERENT},
\newblock (2022), arXiv:2202.11002.

\bibitem{BaBar:2017tiz}
BaBar, J.~P. Lees {\em et~al.}, {Search for Invisible Decays of a Dark Photon
  Produced in ${e}^{+}{e}^{-}$ Collisions at BaBar},
\newblock Phys. Rev. Lett. {\bf 119}, 131804 (2017), arXiv:1702.03327.

\bibitem{Banerjee:2019pds}
D.~Banerjee {\em et~al.}, {Dark matter search in missing energy events with
  NA64},
\newblock Phys. Rev. Lett. {\bf 123}, 121801 (2019), arXiv:1906.00176.

\bibitem{NA64:2017vtt}
NA64, D.~Banerjee {\em et~al.}, {Search for vector mediator of Dark Matter
  production in invisible decay mode},
\newblock Phys. Rev. D {\bf 97}, 072002 (2018), arXiv:1710.00971.

\bibitem{PhysRevLett.125.171802}
SENSEI Collaboration, L.~Barak {\em et~al.}, Sensei: Direct-detection results
  on sub-gev dark matter from a new skipper ccd,
\newblock Phys. Rev. Lett. {\bf 125}, 171802 (2020).

\bibitem{delta_a1}
H.~Banerjee, B.~Dutta, and S.~Roy, {Supersymmetric gauged $
  \mathrm{U}{(1)}_{L_{\mu }-{L}_{\tau }} $ model for electron and muon $(g-2)$
  anomaly},
\newblock JHEP {\bf 03}, 211 (2021), arXiv:2011.05083.

\bibitem{constraint_cos1}
B.~Ahlgren, T.~Ohlsson, and S.~Zhou, Comment on ``is dark matter with
  long-range interactions a solution to all small-scale problems of
  $\ensuremath{\Lambda}$ cold dark matter cosmology?'',
\newblock Phys. Rev. Lett. {\bf 111}, 199001 (2013).

\bibitem{constraint_cos2}
K.~J. Kelly, M.~Sen, W.~Tangarife, and Y.~Zhang, Origin of sterile neutrino
  dark matter via secret neutrino interactions with vector bosons,
\newblock Phys. Rev. D {\bf 101}, 115031 (2020).

\bibitem{constraint_babar}
BaBar collaboration, {BaBar collaboration}, Search for a muonic dark force at
  babar,
\newblock Phys. Rev. D {\bf 94}, 011102 (2016).

\bibitem{constraint_nt}
W.~Altmannshofer, S.~Gori, M.~Pospelov, and I.~Yavin, Neutrino trident
  production: A powerful probe of new physics with neutrino beams,
\newblock Phys. Rev. Lett. {\bf 113}, 091801 (2014).

\bibitem{exp_Borexino_1}
A.~Kamada and H.-B. Yu, Coherent propagation of pev neutrinos and the dip in
  the neutrino spectrum at icecube,
\newblock Phys. Rev. D {\bf 92}, 113004 (2015).

\bibitem{exp_Borexino_2}
S.~Gninenko and D.~Gorbunov, {Refining constraints from Borexino measurements
  on a light Z'-boson coupled to L\ensuremath{\mu}-L\ensuremath{\tau} current},
\newblock Phys. Lett. B {\bf 823}, 136739 (2021), arXiv:2007.16098.

\bibitem{exp_ATLAS_1}
ATLAS Collaboration, {ATLAS Collaboration}, Measurements of four-lepton
  production at the $z$ resonance in $pp$ collisions at $\sqrt{s}=7$ and 8 tev
  with atlas,
\newblock Phys. Rev. Lett. {\bf 112}, 231806 (2014).

\bibitem{exp_ATLAS_2}
W.~Altmannshofer, S.~Gori, S.~Profumo, and F.~S. Queiroz, {Explaining dark
  matter and B decay anomalies with an $L_\mu - L_\tau$ model},
\newblock JHEP {\bf 12}, 106 (2016), arXiv:1609.04026.

\bibitem{exp_CMS}
CMS, A.~M. Sirunyan {\em et~al.}, {Search for an $L_{\mu}-L_{\tau}$ gauge boson
  using Z$\to4\mu$ events in proton-proton collisions at $\sqrt{s} =$ 13 TeV},
\newblock Phys. Lett. B {\bf 792}, 345 (2019), arXiv:1808.03684.

\end{thebibliography}

\end{document}